\documentclass[twocolumn,astrosymb]{aastex631}
\pdfoutput=1
\usepackage{rotating}
\usepackage{booktabs}
\usepackage{textcomp, gensymb}
\usepackage{tabularx}
\usepackage{hyperref}

\shorttitle{Spot Filling Factors and Stellar Rotation}
\shortauthors{P\'erez Paolino et al.}
\graphicspath{{./}{figures/}}

\begin{document}

\title{Correlating Changes in Spot Filling Factors with Stellar Rotation: The Case of LkCa 4}

\correspondingauthor{Jeffrey Bary}
\email{jbary@colgate.edu}

\author[0000-0002-4128-7867]{Facundo P\'erez Paolino}
\affiliation{Colgate University, 13 Oak Drive, Hamilton, NY 13346, USA}

\author[0000-0001-8642-5867]{Jeffrey S. Bary}
\affiliation{Colgate University, 13 Oak Drive, Hamilton, NY 13346, USA}
\affiliation{Visiting astronomer at the Infrared Telescope Facility}
\altaffiliation{NASA's Infrared Telescope Facility is operated by the University of Hawaii under contract 80HQTR19D0030 with the National Aeronautics and Space Administration.}


\author[0000-0003-1517-3935]{Michael S. Petersen}
\affiliation{Visiting Astronomer at the Infrared Telescope Facility}
\affiliation{Institute for Astronomy, University of Edinburgh, Royal Observatory, Blackford Hill, Edinburgh EH9 3HJ, UK}

\author[0000-0002-4479-8291]{Kimberly Ward-Duong}
\affiliation{Visiting Astronomer at the Infrared Telescope Facility}
\affiliation{Department of Astronomy, Smith College, 10 Elm Street, Northampton, MA 01063, USA}

\author[0000-0003-2053-0749]{Benjamin M. Tofflemire}
\altaffiliation{51 Pegasi b Fellow}
\affiliation{Department of Astronomy, The University of Texas at Austin, Austin, TX 78712, USA}

\author[0000-0002-7821-0695]{Katherine B. Follette}
\affiliation{Department of Physics and Astronomy, Amherst College, Amherst, MA 01003, USA}

\author{Heidi Mach}
\affiliation{Allegheny College, 520 N. Main Street Meadville, PA 16335, USA}



\begin{abstract}

We present a multi-epoch spectroscopic study of LkCa~4, a heavily spotted non-accreting T~Tauri star. Using SpeX at NASA's Infrared Telescope Facility (IRTF), 12 spectra were collected over five consecutive nights, spanning $\approx$~1.5 stellar rotations. Using the IRTF SpeX Spectral Library, we constructed empirical composite models of spotted stars by combining a warmer (photosphere) standard star spectrum with a cooler (spot) standard weighted by the spot filling factor, $f_{spot}$. The best-fit models spanned two photospheric component temperatures, $T_{phot}$~=~4100~K (K7V) and 4400~K (K5V), and one spot component temperature, $T_{spot}$~=~3060~K (M5V) with an $A_V$ of 0.3. We find values of $f_{spot}$ to vary between 0.77 and 0.94 with an average uncertainty of $\sim$0.04. The variability of $f_{spot}$ is periodic and correlates with its 3.374~day rotational period. Using a mean value for $f^{mean}_{spot}$ to represent the \textit{total} spot coverage, we calculated spot corrected values for $T_{eff}$ and $L_\star$. Placing these values alongside evolutionary models developed for heavily spotted young stars, we infer mass and age ranges of 0.45-0.6~$M_\odot$ and 0.50-1.25~Myr, respectively. These inferred values represent a twofold increase in the mass and a twofold decrease in the age as compared to standard evolutionary models. Such a result highlights the need for constraining the contributions of cool and warm regions of young stellar atmospheres when estimating $T_{eff}$ and $L_\star$ to infer masses and ages as well as the necessity for models to account for the effects of these regions on the early evolution of low-mass stars.

\end{abstract}



\section{Introduction} \label{sec:intro}

Accurate age determinations for pre-main-sequence (PMS) stars are essential to understanding the formation and evolution of stars and planetary systems. One popular method for constraining the ages of PMS stars relies on a direct comparison between the observed stellar luminosities and surface temperatures and those predicted by theoretical evolutionary models \citep[e.g.,][]{DAntona1994,Soderblom2014,Baraffe2015}. Unfortunately, young stars are complex systems often characterized by strong magnetic fields, rapid rotation rates, excess emission from circumstellar material, mass accretion onto the stellar surfaces, and mass outflow from disk and stellar winds \citep[e.g.,][]{Hartmann2016}. Chromospheric and coronal activity are heightened, leading to strong flares producing large fluxes of high-energy photons \citep[e.g.,][]{feig1999,petrov2011}. As part of the heightened activity, the large-scale inhibition of convection by strong magnetic fields in these systems is quite possible, resulting in the formation of starspots covering significant fractions of the stellar surfaces. Given the episodic or transient nature of these phenomena, young stellar systems exhibit variability on timescales as short as hours across all wavelengths. Such activity often limits our ability to constrain otherwise straightforward observable stellar parameters. Ages inferred from comparisons of effective temperatures and stellar luminosities to those predicted by standard evolutionary models \citep[e.g.,][]{Baraffe2015} typically result in large spreads in the ages of stars residing in the same cluster. While some of this spread may be due to different star formation epochs that have occurred within the same region, ignoring the effects of spots on the observable quantities and on stellar evolution likely contributes to this spread, confusing our understanding of the star forming history. Therefore, the presence of large starspots on the surfaces of a sizable fraction of young stars and the lack of evolutionary models that account for spots are potentially responsible for some of the spread in the ages, masses, and evolutionary statuses inferred for stars in a given young cluster \citep{Preibisch2012,Soderblom2014}.

Large complexes of cool spots rotating with the surfaces of low-mass PMS stars produce periodic variability, which can be used profitably to measure rotation periods for these objects \citep[e.g.,][]{bouvier1995,Herbst2007,Grankin2008}. Such spots also provide a reasonable explanation for the systematic color anomalies and optical/infrared spectral type mismatches observed for T~Tauri stars, both accreting and non-accreting \citep{Gullbring1998,Vacca2011,Debes2013,Bary2014,czekala2015,Kastner2015,Gully2017}. 

\citet{Debes2013} and \citet{Bary2014} demonstrate that the near-infrared (NIR) spectra of TW~Hya and DQ~Tau, respectively, can be modeled with empirical composite spectra made from a weighted average of two standard star spectra -- a warmer standard representing the photosphere and a cooler one representing the spot. In both cases, the authors find that cool spots may cover over 50\% of the surfaces of the stars. \citet{Donati2014} use multi-epoch spectropolarimetric observations (R$\sim$65,000) of LkCa~4, a weak-line T~Tauri star and the subject of the study presented herein, to construct tomographic maps of the stellar surface and to study the magnetic topology of the star. Their results indicate the presence of large cool spots covering more than 20\% of the surface as well as the existence of large warm plages. \citet{herczeg2014} use TiO features in the near-IR to revise the spectral type of LkCa~4 from a K7 to a later M1.5, likely highlighting the effect of spots on single-band temperature measurements. \citet{Gully2017} (from here on GS17) also observe LkCa~4 at high-spectral resolution (R$\sim$45,000) in the near-IR with IGRINS. Applying a two-temperature atmospheric model to fit their data as well as the TiO bands in the spectra of \citet{Donati2014}, GS17 establish the presence of a large spot or spot complex that covers nearly 80\% of the stellar surface. The surprisingly large discrepancy between filling factors determined by \citet{Donati2014} and GS17 can be reconciled by the fact that the Zeeman Doppler Imaging (ZDI) technique employed by Donati et al.\ is insensitive to collections of smaller spots.

Using high-resolution iSHELL spectra (R$\sim$47,000), \citet{Flores2019,Flores2022} measure magnetic field strengths on the surfaces of T~Tauri stars and correlate magnetic field strengths to measurements of stellar temperatures. The results of Flores et al.\ indicate that a correlation likely exists between spots and effective temperatures of these sources.

Collectively, these studies highlight the uncertainty and complexity that starspots introduce when using evolutionary models to infer ages and masses of highly active PMS stars. They also illustrate the importance of developing new evolutionary models of spotted stars, incorporating the physical mechanisms that produce the spots as well as predicting their impact on the evolution of young stars \citep[e.g.,][]{Feiden2016,Somers2020}. Such models will improve our efforts to confidently and accurately infer the ages of PMS stars and the clusters in which they form. Constraining these models will require a simple and direct method for determining spot filling factors and spot temperatures for large samples of PMS stars across the mass spectrum.

Toward this goal, we present a multi-epoch, medium-resolution, NIR spectroscopic study of LkCa~4 in which we constrain spot sizes and temperatures and correlate the changes in the spot filling factors with the rotational phase of the star. Using our best-fit model parameters for photospheric temperature ($T_{phot}$), spot temperature ($T_{spot}$), and spot filling factors ($f_{spot}$), we are able to reproduce the $V$-band variability observed during a time frame that overlaps with our spectral observations. We show how our results compare with the studies mentioned above, which were conducted at much higher spectral resolution \citep{Donati2014,Gully2017}. The observations presented benefit from consistent temporal coverage over five consecutive nights or roughly 1.5 stellar rotations. Although the absolute value of the total spot coverage depends on the model-dependent photosphere and spot temperatures, the temporal coverage permits us to better constrain the total fraction of the stellar surface covered by spots than that of a single observation.

First, we outline the observations and calibration steps in Section \ref{sec:obs}. In Sections \ref{sec:starspot} \& \ref{sec:fitting}, we describe the empirical composite spectral models and determine the best-fit filling factors and temperature ranges for the photosphere and the spots. In Section \ref{sec:ffactorallnights}, we present the strong correlation we find between the photometric variability and the changes observed in the spot filling factors suggesting our observations and spectral models are sensitive to the rotation of the star. In Section~\ref{CO}, we take a small digression to explore another spectral type indicator, the $D_{\text{CO}}$ spectral index based on the 2.29~\micron\ CO bandhead \citep{marmolqueralto} to test its sensitivity to spots and its consistency with other temperature indicators. In Section \ref{sec:spotsmodels}, we use the best-fit model parameters to calculate spot-corrected $T_{eff}$ and $L_\star$. We then place these spot-corrected values on the HR~diagram alongside evolutionary tracks and isochrones predicted by both standard and spotted star evolutionary models \citep{Baraffe2015,Somers2020} and discuss the results.
\vspace{0cm}

\section{Observations} \label{sec:obs}

We observed LkCa~4 using SpeX, a medium-resolution cross-dispersed NIR spectrograph at NASA's Infrared Telescope Facility (IRTF) atop Maunakea over five consecutive nights on UT January 6-10 2019. Using the short-wavelength cross-dispersed mode \citep[SXD;][]{rayner2003} with the 0.\arcsec3 $\times$ 15\arcsec~slit (R$\sim$2000), we collected a total of $12$ spectra of the target as part of a larger program to study starspots and accretion activity in PMS systems. The SXD setting provides continuous wavelength coverage from 0.7-2.55~$\micron$. The goals of this project were achieved due to favorable weather conditions, which permitted consistent monitoring of more than one full rotation of LkCa~4.

\begin{figure}
\includegraphics[width=8.5cm, height=10cm]{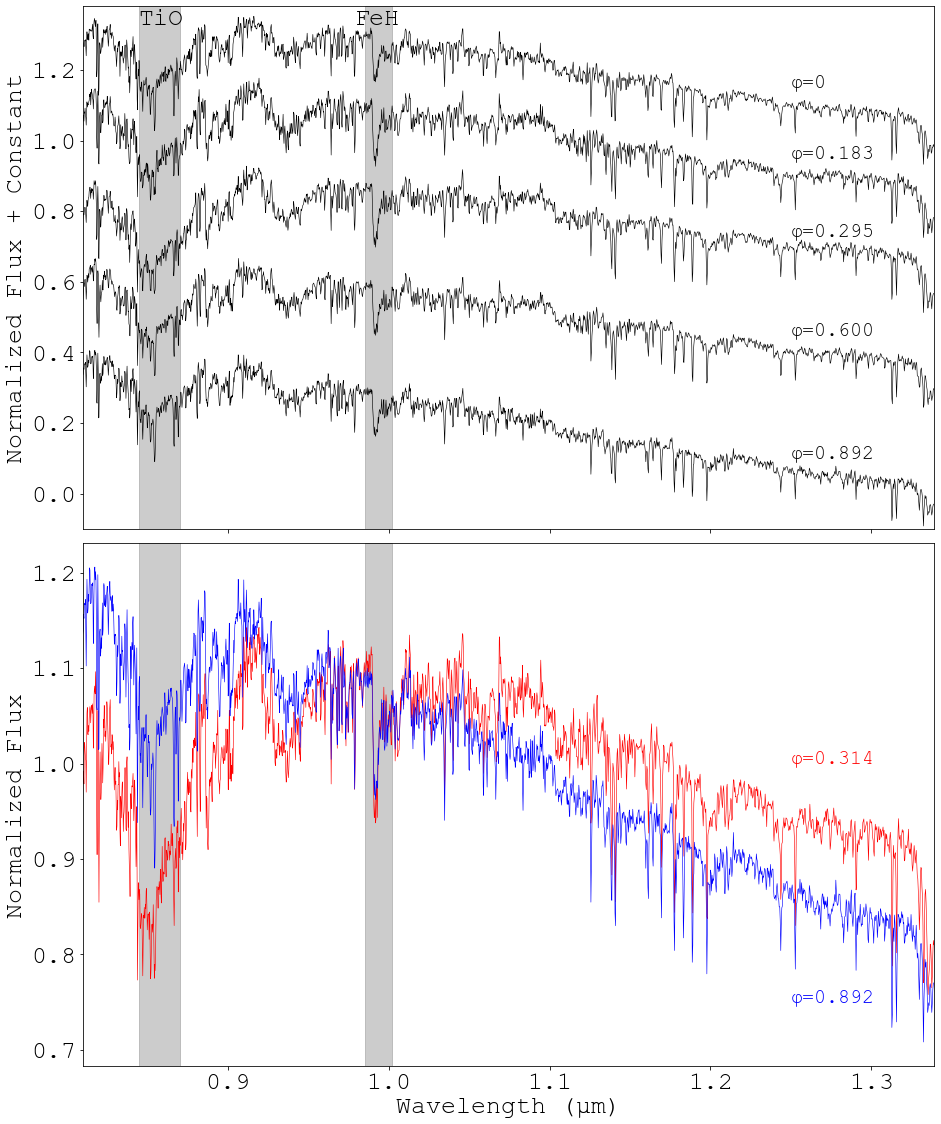}
\caption{(Top) Five partial LkCa~4 SpeX spectra are plotted from 0.8-1.35~$\micron$; one from each night of the observing run. The spectra have been normalized by their mean flux values in the plotted wavelength range and shifted by adding a constant for display purposes. Changes in the strengths of TiO and FeH absorption bands are apparent, as well as variations in the overall spectral shapes. Gray-shaded areas highlight the spot-sensitive TiO and FeH features. (Bottom) Two LkCa~4 spectra plotted at maximum (red) and minimum (blue) excursions from the average spectrum. \label{fig:allnightspec}}
\end{figure}

The data were collected using an AB nod sequence typical of long-slit near-IR spectra acquisition. The subtraction of the 2D spectral image pairs allows for the efficient removal of terrestrial OH emission lines, background, and dark current. A0V telluric standards HD~27761 and HD~24000 were observed close in airmass ($\Delta \sec z~\leq0.1$) to the target and were used to remove telluric absorption features and to calibrate the target spectra. Flat-field corrections, wavelength calibrations, spectral extraction, co-adding, telluric corrections, and merging of the spectral orders were performed using SpexTool v4.0.5, an IDL-based reduction package described by \citet{cushing2004}. Details of the observations can be found in Table~\ref{tab:obsdata}. Sample spectra covering the 0.8 to 1.35~\micron\ region are shown in Figure~\ref{fig:allnightspec}, highlighting the nature of the variability observed in the shorter wavelength regions of the spectra.

\begin{deluxetable}{rhcccc}
\tabletypesize{\footnotesize}
\setlength{\tabcolsep}{5pt}
\tablecaption{LkCa 4 SpeX Observations\label{tab:obsdata}}
\tablehead{
\colhead{UT Date Time} & \nocolhead{Phase} & \colhead{Airmass} & \colhead{Exposure time} & \colhead{Telluric} & \colhead{$\Delta \text{sec\ z}$}\\
\colhead{} & \nocolhead{($\phi$)}& \colhead{(sec z)} & \colhead{(sec~$\times$~$n$ coadds)} & \colhead{Standard}}

\decimals
\startdata
Jan 6 05:07:28  & 0.000 & 1.222 & 75~$\times$~8   & HD 24000 & 0.10\\
Jan 6 07:24:35  & 0.028 & 1.012 & \nodata         & HD 24000 & 0.01\\
Jan 6 09:29:46  & 0.054 & 1.122 & \nodata         & HD 27761 & 0.03\\
Jan 7 04:59:00  & 0.295 & 1.237 & \nodata         & HD 24000 & 0.11\\
Jan 7 06:35:23  & 0.314 & 1.038 & \nodata         & HD 24000 & 0.02\\
Jan 8 05:43:50  & 0.600 & 1.108 & \nodata         & HD 24000 & 0.06\\
Jan 8 07:19:17  & 0.620 & 1.011 & \nodata         & HD 27761 & 0.01\\
Jan 8 10:06:01  & 0.654 & 1.246 & \nodata         & HD 27761 & 0.05\\
Jan 9 05:19:47  & 0.892 & 1.153 & \nodata         & HD 24000 & 0.07\\
Jan 9 07:07:34  & 0.914 & 1.013 & 150~$\times$~4  & HD 27761 & 0.01\\
Jan 10 04:55:55 & 0.183 & 1.209 & \nodata         & HD 24000 & 0.10\\
Jan 10 07:02:14 & 0.209 & 1.014 & \nodata         & HD 27761 & 0.01\\
\enddata
\end{deluxetable}

\section{Empirical Models of Spotted Stars} \label{sec:starspot}

We constructed two-temperature models of spotted stars as empirical composite spectra following the same procedure outlined in \citet{Debes2013} and \citet{Bary2014}. We use the term \textit{empirical composite} to clearly indicate that these model spectra are not generated from synthetic stellar atmospheric models. Instead, they are produced using the spectra of standards found in the SpeX IRTF Library \citep{rayner2009, Cushing2005}. In the two-temperature models, the spectrum of the warmer standard represents the photosphere, $F_{\lambda}$($T_{phot}$), and the cooler standard represents the spots, $F_{\lambda}$($T_{spot}$). Therefore, changing model parameters $T_{phot}$ or $T_{spot}$ is achieved by selecting different spectral standards from the SpeX Library. 
The composite spectral models are generated using the following

\begin{equation}
F_{\lambda, model} = F_{\lambda}(T_{phot})(1-f_{spot}) + F_{\lambda}(T_{spot})f_{spot} C_{bb}
\end{equation}

\noindent where $F_{\lambda}$($T_{phot}$) and $F_{\lambda}$($T_{spot}$) are normalized template spectra, $f_{spot}$ is the \textit{instantaneous} spot filling factor (i.e., the fraction of the observable stellar surface that is covered by spots following the convention adopted by GS17), and $C_{bb}$ is a scaling constant defined as the ratio of the Planck functions of the photosphere and the spot at 1.1~$\micron$. $C_{bb}$ approximates the relative normalized flux units of the two spectra based on their effective temperatures. Before combining, the wavelength arrays of the standard star spectra are aligned through a one-dimensional interpolation using the \textit{interp1d} algorithm found in SciPy \citep{jones2014}. The resulting composite model spectrum is renormalized before fitting to the target spectra.

We have chosen to use dwarf spectral standards when constructing the models similar to \citet{Debes2013} and 
\citet{Bary2014}. The spectral standards used to construct the spotted star models as well as the spectral types, 
effective temperatures, variability status, and $B-V$ color excesses for each source are listed in 
Table~\ref{tab:stardata}. We note that M~dwarfs are well known for their aperiodic variability due to strong 
stochastic flares \citep{Hartman2011}. We find that six of the M~dwarf standards selected to represent the spots 
are identified as eruptive variables. However, we proceed with using the M~dwarfs as standards assuming that the 
short-lived nature of the eruptions is not likely to impact the single-epoch observations presented in the SpeX 
library. \citet{rayner2009} measure color excesses for the library stars and do not present dereddened spectra for 
stars with $E(B-V)<0.108$. The color excesses quoted for Gl~406 (M6V) and Gl~466C (M7V) are significant. Therefore, 
these standard star spectra were dereddened with a standard interstellar extinction law \citep{Fitzpatrick1999} 
prior to using them to construct spectral models.

In general, optical and infrared TiO and FeH features are temperature sensitive and are considered good, yet 
complicated spot indicators \citep{herbst1990,Neff1995,oneal1996,schiavon1997}. Therefore, it is important to note 
different sensitivities between TiO and FeH that may impact the individual constraints they place on the best-fit 
composite spectra and values for $f_{spot}$. For instance, the FeH Wing-Ford band at 0.99~$\micron$ is sensitive to 
changes in surface gravity with the feature appearing to be strongest in the spectra of the coolest dwarf stars 
\citep{schiavon1997}. In fact, its presence and strength in NIR spectra of unresolved stellar populations have 
frequently been used to measure the contribution from cool, dwarf stars \citep[e.g.,][]
{Couture1993,schiavon1997,cenarro2003}. However, with regards to its sensitivity to small changes in $\log$~g, 
\citet{Bary2014} found little difference in the strengths of FeH between synthetic spectra of a K5IV with 
$\log$~g~=~3.5, representing a ``puffy" T~Tauri star, and a K5V dwarf star with $\log$~g~=~4.5 
\citep{coelho2005A&A}. Such a similarity likely indicates that larger differences in the surface gravity are 
required to produce an effect on the FeH band strengths to be detectable with moderate resolution spectroscopy.

In addition to surface gravity effects, the magnetic sensitivities of both TiO and FeH molecular states and 
transitions to Zeeman effects is also an important consideration. Absorption features associated with both bands 
have been shown to be quite sensitive to magnetic fields and have been used to measure magnetic field strengths on 
K- and M-type stars \citep[e.g.,][]{afram2015}. FeH has gained considerable attention as a probe of magnetic field 
strengths on M dwarfs, which are too cool to possess strong, magnetically sensitive atomic features 
\citep{valenti2001,reiners2006,shulyak2014,afram2019,kochukhov2021}. 

Given that the composite spectra are constructed with template spectra of M dwarfs representing the cooler spotted 
regions that likely possess magnetic fields that are stronger than the non-spot regions, it is important to 
acknowledge that the M dwarf templates possess TiO and FeH features that are affected by strong magnetic fields. 
For instance, \citet{afram2019} find a range of 3-6~kG fields with an average of 5~kG for a sample of nine M1-7 
dwarfs. \citet{shulyak2019} measure magnetic fields for a larger sample of 29 active M dwarfs and similarly find 
field strengths in the 1-7~kG range. On average, these M dwarfs have greater field strengths than the 0.71-3.24~kG 
range measured by \citet{Flores2022} for a sample of 40 K and M spectral type T~Tauris stars. The TiO and FeH 
contributions to the composite spectra from the M dwarf templates will likely incorporate some effects of a 
magnetic field within the 1-7 kG range.

\begin{deluxetable}{lcccc}
\setlength{\tabcolsep}{3pt}
\tablecaption{SpeX Spectral Standards \label{tab:stardata}}
\tablehead{
\colhead{Star} & \colhead{Spectral Type} & \colhead{$T_{eff}$\tablenotemark{a}} & \colhead{Variable} & 
\colhead{$E(B-V)$\tablenotemark{b}}
}
\decimals
\startdata
HD 45977   & K4V$_p$ & 4600 & none    & 0.012\\
HD 36003   & K5V$_p$ & 4400 & none    & -0.037\\
HD 237903  & K7V$_p$ & 4100 & none    & 0.022\\
HD 19305   & M0V$_p$ & 3850 & none    & $<$0.019\\
\hline
HD 42581   & M1V$_s$ & 3660 & UV\tablenotemark{c} & $<$0.018\\
Gl 806     & M2V$_s$ & 3560 & var\tablenotemark{d} & \nodata\\
Gl 388     & M3V$_s$ & 3430 & UV\tablenotemark{e} & $<$0.009\\
Gl 213     & M4V$_s$ & 3210 & BY\tablenotemark{f} & $<$0.009\\
Gl 51      & M5V$_s$ & 3060 & UV\tablenotemark{e} & \nodata\\
Gl 406     & M6V$_s$ & 2810 & UV\tablenotemark{f} & 0.063\\
Gl 644C    & M7V$_s$ & 2680 & UV\tablenotemark{c} & 0.100\\
LP 412-31  & M8V$_s$ & 2570 & UV\tablenotemark{g} & \nodata\\
\enddata
\tablenotetext{a}{All temperatures are from \citet{Pecaut2013}.}
\tablenotetext{b}{Color excesses reported in \citet{rayner2009}.}
\tablenotetext{c}{\citet{Gershberg1999}}
\tablenotetext{d}{Uncharacterized variability.\citep{Alfonso-Garz2012}}
\tablenotetext{e}{\citet{Jones2016}}
\tablenotetext{f}{\citet{samus2017}}
\tablenotetext{g}{\citet{stelzer2006}}
\end{deluxetable}
\vspace{0.0cm}

Previous studies of spotted T~Tauri stars have indicated that the spectral types representing $T_{phot}$ in two-temperature models will be similar to the optically derived spectral types for the stars 
\citep{Debes2013,herczeg2014,Bary2014,Gully2017}. Therefore, the spectral templates chosen to represent 
$F_\lambda$($T_{phot}$) in our composite models bracket the K7V spectral type reported for LkCa~4 in the literature 
\citep{Herbig1986, Strom1994, Edwards1995, White2001, Grankin2013}. We constrained the parameter 
$F_\lambda$($T_{phot}$) by selecting four photospheric templates: M0V$_p$, K7V$_p$, K5V$_p$, and K4V$_p$ 
(3850~K~$\leq$~$T_{phot}$~$\leq$~4600~K). The spectral types of the templates representing $F_\lambda$($T_{spot}$) 
were confined to the range between M8V$_s$ and M1V$_s$ (2570~K~$\leq$~$T_{spot}$~$\leq$~3660~K). This range of spot 
temperatures encompasses the values suggested by previous studies (i.e., GS17) and fits the typical spot-to-photosphere temperature ratios \citep{Strassmeier2009,Fang2018}. The contribution of the spot template to the 
composite spectrum is weighted by the spot filling factor, $f_{spot}$, which we allow to vary from 0.0 to 1.0. It 
is important to note that cooler spots with smaller filling factors will mimic warmer spots with larger filling 
factors leading to an inherent degeneracy in these models.

\subsection{Model Fitting} \label{sec:fitting}
We searched for the best-fit model for each of the 12 epochs of LkCa~4 spectra by varying four model parameters: 
$T_{phot}$, $T_{spot}$, $f_{spot}$, and $A_V$. The values for $T_{phot}$, $T_{spot}$ and $f_{spot}$ were 
constrained by the parameter space defined above in Section~\ref{sec:starspot}. We let $A_V$ vary between 0.0 and 
1.0 in steps of 0.1 to encompass the two values of 0.35 \citep{Gully2017} and 0.69 \citep{Kenyon1995} reported in 
the literature. The LkCa~4 spectra were dereddened with the same standard interstellar extinction law used for the 
M~dwarf standards.

We selected three spectral windows to constrain the best-fit models. The first is a 
broad spectral window stretching from 0.8\footnote{The spectra in the SpeX Library have 
a short-wavelength cutoff at 0.8~$\micron$ because they were collected prior to the 
2014 SpeX upgrade to a Hawaii-2RG detector, which pushed the sensitivity down to 
0.7~$\micron$.} to 1.35~\micron, which we will designate $F_{0.8-1.35~\micron}$. The 
large wavelength coverage of this window will force the best-fit models to accurately 
reproduce the shape of the continuum in a region of the spectrum that is most 
significantly affected by interstellar reddening. The other two spectral windows center 
on two spot-sensitive molecular absorption features: TiO band ($\lambda$=0.845$\-$
0.870~$\micron$) and the Wing-Ford FeH band ($\lambda$=0.985\-1.02~$\micron$). We note that 
both of the molecular features are included within the larger spectral window. Given 
the comparatively large wavelength coverage of the $F_{0.8-1.35~\micron}$ removing 
either or both of the features from the window while performing the fitting routine 
(see below) affects the $\chi^2_{\rm red}$ values \footnote{All $\chi^2$ values 
presented in the paper are reduced $\chi^2$ values regardless of the subscript.} 
on the order of 0.01\%. Therefore, we simply perform the fits to this window without 
excluding the TiO and FeH absorption features. We will refer to these three spectral 
regions as spot indicators or indicators as shorthand.

An initial round of fits between the composite models and the target data were 
performed using the Levenberg-Marquardt minimization algorithm found in \textit{lmfit} 
\citep{Newville2014}. \textbf{The $\chi^2_{\rm red}$ values for these fits are systematically 
large due to the noise in the composite models and the consistently poor fits to the atomic 
absorption features.} The $\chi^2_{\rm red}$ values did not exactly follow a normal distribution. 
The two-temperature models with $f_{spot}$ that possessed the lowest $\chi^2_{\rm red}$ values 
were then used as the starting point for the walkers in an \textit{emcee} Markov Chain Monte Carlo sampler \citep{Foreman-Mackey2013}. The MCMC procedure probed the posterior probability 
density function of $f_{spot}$. In practice, the posterior probability density functions 
resembled normal distributions with maxima that closely corresponded to the models with the 
lowest $\chi^2_{\rm red}$ values. We have adopted the 1$\sigma$ widths of these distributions 
as the uncertainties in $f_{spot}$. This was performed for every observation and all 
parameter space, yielding the uncertainties in $f_{spot}$. 

For the three different spot indicators, we find good agreement between the best-fit 
model parameters, $T_{phot}$, $T_{spot}$, and $f_{spot}$ constrained by the
$F_{0.8-1.35~\micron}$ spectral window and TiO feature. 
 
\begin{figure*}
\centering
\includegraphics[height=21cm]{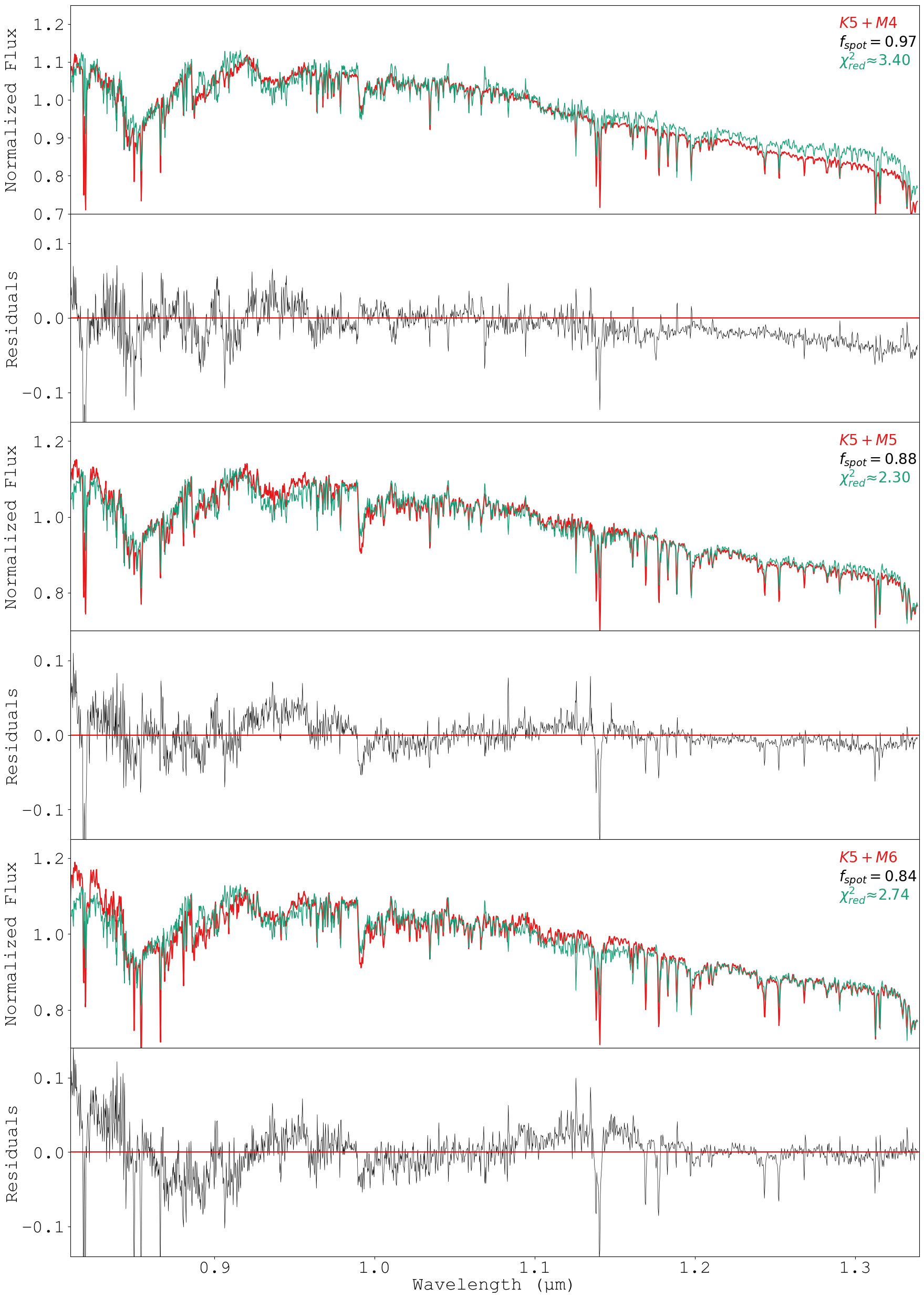}
\caption{(Top) A dereddened LkCa~4 spectrum (green; $\phi=0.000$) compared to the best-fit empirical composite spectra K5V$_p$+M4V$_s$ (red). Beneath are residuals plotted with respect to zero (red line). (Middle) and (Bottom) These plots are similar comparisons for the best fits using K5V$_p$+M5V$_s$ and K5V$_p$+M6V$_s$ models, respectively. The $\chi^2_{red}$ values were calculated in the $0.8-1.35~\micron$ region.}
\label{fig:lkca4fit}
\end{figure*}

In Figure~\ref{fig:lkca4fit}, we graphically illustrate the goodness of fit by 
presenting comparisons of one LkCa~4 spectrum to three composite models with different 
$f_{spot}$ values in the 0.8-1.35~$\mu$m window. By visual inspection, the
K5V$_p$+M5V$_s$ model with $f_{spot}$~=~0.88 and $\chi^2_{\rm red}$~$\approx 2.3$ is 
an overall better fit than the K5V$_p$+M4V$_s$ and K5V$_p$+M6V$_s$ models. For a 
difference in $\chi^2_{\rm red}$ of $\sim$1.1 and 0.44, significant differences in the 
residuals and the quality of the fits to the strengths of the TiO features and the 
general shapes of the 0.8-1.35~$\mu$m region of the spectra are evident. Similar to 
other two-temperature models presented in \citet{Bary2014}, atomic absorption features 
are not well fit and contribute to the large $\chi^2_{\rm red}$ as compared to the fits 
to the TiO or FeH features. Similar differences between the goodness of fit for the 
K7V$_p$+M4V$_s$, K7V$_p$+M5V$_s$, and K7V$_p$+M6V$_s$ models are observed 
with the K7V$_p$+M5V$_s$ producing the lowest $\chi^2$ value.

\vspace{0cm}

\begin{figure*}
\centering
\includegraphics[height=13cm]{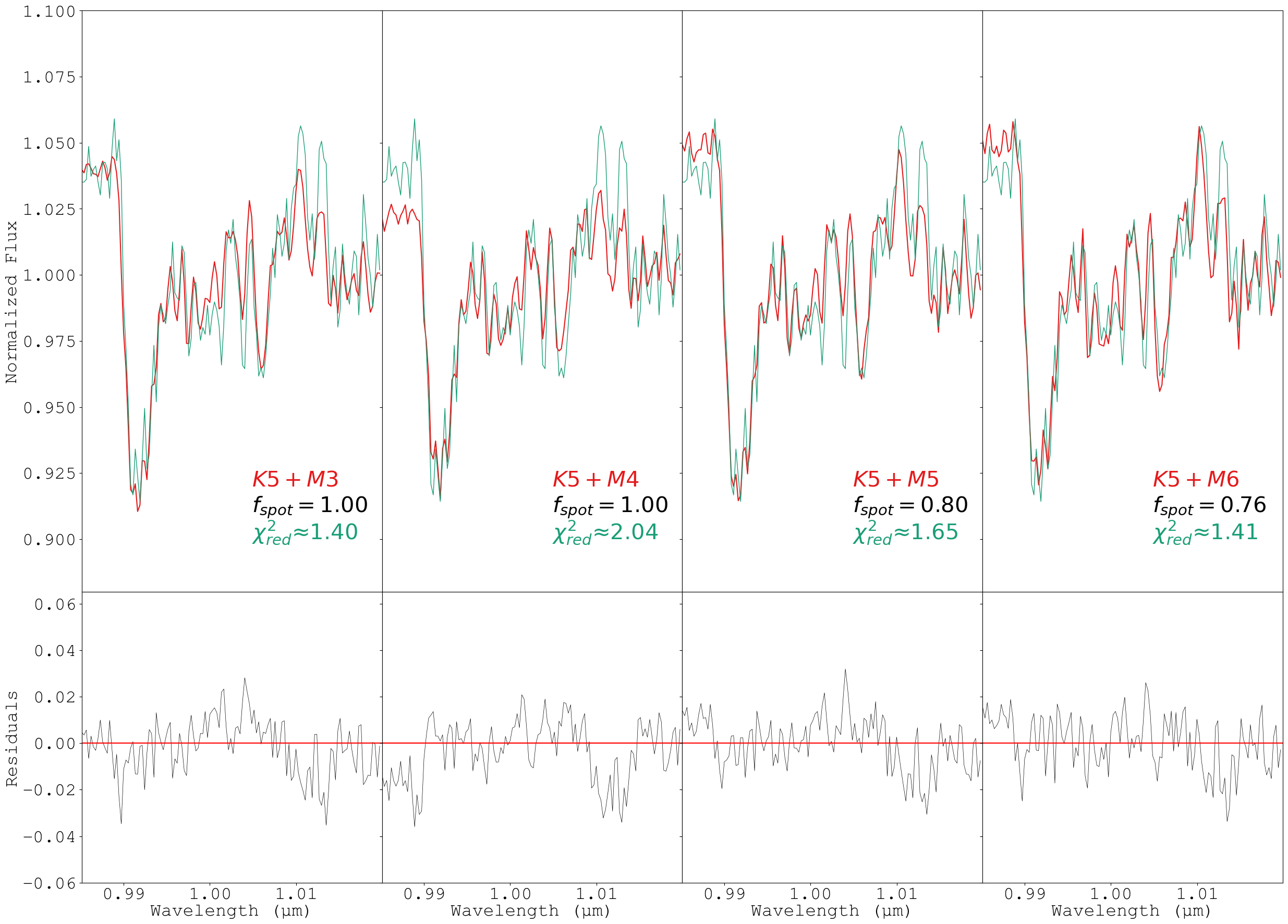}
\caption{(Top left) A dereddened LkCa~4 spectrum (green; $\phi=0.000$) and 
the best-fit empirical composite spectrum with K5V$_p$+M3V$_s$ (red) 
plotted over the wavelengths corresponding to the FeH window 
($l\lambda$~=~0.985-1.02$\micron$). (Bottom left) Plotted are the residuals of the fit. (Top middle, top right, and far right) Similar plots comparing the best fits to the FeH feature for ther K5V$_p$+M4V$_s$, K5V$_p$+M5V$_s$ and K5V$_p$+M6V$_s$ models, respectively.}
\label{fig:fehfits}
\end{figure*}

It is apparent in Figure~\ref{fig:lkca4fit}, that the models do not fit the FeH 
feature in a predictable manner or one that is consistent with the other spot 
indicators. On average over the 12 epochs of observations, the FeH fits point to three 
possible best-fit models with considerably different values for $f_{spot}$. For the 
K5V$_p$+M3V$_s$ model, the value of $f_{spot}$ is 1.0 essentially selecting an M3V 
spectrum as the best fit to all 12 epochs. By contrast, the K5V$_p$+M5V$_s$ and 
K5V$_p$+M6V$_s$ models were equally good fits with $f_{spot}$ values falling in the 
range of 0.72-0.85 over the 12 epochs. Similar behavior of the FeH fits was 
observed for the models in which the K5V was replaced with a K7V standard. The spot 
filling factors decreased as expected for a model with a cooler photosphere. To 
illustrate the goodness of fit to the FeH feature, we present a similar plot for the FeH
comparing four model fits in  Figure~\ref{fig:fehfits}. The K5V$_p$+M4V$_s$ fits are the 
poorest due to the mismatch between the model and the data between 0.985 and 
0.990~$\micron$. The other three models do a far better job of fitting these short 
wavelengths and differ mostly in the way they fits the small components of the feature. 
The similarities between the other three fits and the outlier nature of the 
K5V$_p$+M4V$_s$ points to a potential problem with the M4V standard. If this is the 
case, then the FeH feature may fit all four of these models equally well over a range 
of filling factors rendering it a less useful spot indicator than $F_{0.8-1.35~\micron}$
and TiO. Therefore, we will treat the best-fit models based on these spot indicators
as the most reliable, but will include model parameters associated with the FeH indicator
in the following discussion where we believe it is useful and instructive.

\vspace{0cm}

In Table~\ref{tab:spotfitdata}, we present the values for $f_{spot}$ values 
associated with the best-fit models constrained by the TiO and $F_{0.8-1.35~\micron}$. 
The values listed for each of the 12 observations correspond to two photospheric 
templates, $F_{\lambda}$(K5V$_{p}$) and $F_{\lambda}$(K7V$_{p}$), combined with one 
spot template, $F_{\lambda}$(M5V$_{s}$). The $f_{spot}$ values listed for FeH 
correspond to the same $F(T_{phot})$ and $F(T_{spot})$ and do not represent the model 
with the minimum $\chi^2$ values obtained using this indicator. In the Appendix, Tables~
\ref{tab:wholefittingK5} and \ref{tab:wholefittingK7} we provide a more complete listing
of the best-fit $f_{spot}$ values over a larger range of spot temperatures for all three
spot indicators.

Why do the spot indicators potentially point toward significantly different $f_{spot}$
values for a given pair of $T_{phot}$ and $T_{spot}$? The most straightforward answer seems
to be that the moderate resolution spectra of the FeH feature does not distinguish between
the two-temperature models as well as the TiO feature and the large spectral window. We
also suspect that the differences in the dissociation energies of TiO and FeH
\cite[$D_{\text{TiO}}$~=~7.26~eV; $D_{\text{FeH}}$~=~2.9~eV;][respectively]{Wahlbeck1967,Wang1996}
may contribute to this discrepancy. For instance, FeH will not form within the photosphere
of a K-type star and will only exist within the cooler spot regions. In addition, there may be
some differences in magnetic sensitivity as well as the strength of the magnetic fields
impacting the line strengths of these molecules. Again with the FeH feature being produced
predominantly in the spotted regions and the TiO features formed both within and outside of
the spot, we speculate that the relative line strengths of these features to be more complex
than can be described by our relatively simple two-temperature models that lack any specific
magnetic field parameters. Finally, as described above, the unknown characteristics of the
standard stars given their variability and surface magnetic activity may also affect the
composite models.

\begin{deluxetable*}{c|ccc|ccc}
\tabletypesize{\footnotesize}
\setlength{\tabcolsep}{5pt}
\tablecaption{$f_{spot}$ Values for K5V$_{p}$+M5V$_{s}$ and K7V$_{p}$+M5V$_{s}$ Models \label{tab:spotfitdata}}
\tablehead{
\colhead{Phase} & \multicolumn{3}{c}{$f_{spot}$(K5V$_{p}$+M5V$_{s}$)} &  \multicolumn{3}{c}{$f_{spot}$(K7V$_p$+M5V$_s$)}\\ 
\colhead{($\phi$)} & \colhead{$F_{0.8-1.35\ \micron}$} & \colhead{TiO} & \colhead{FeH}  & \colhead{$F_{0.8-1.35\ \micron}$} & \colhead{TiO} & \colhead{FeH}}
\decimals
\startdata
0.000 & $0.88\pm{0.04}$ & $0.86\pm{0.01}$ & $0.80\pm{0.02}$ & $0.85\pm{0.04}$ & $0.83\pm{0.02}$ & $0.68\pm{0.03}$ \\
0.028 & $0.86\pm{0.04}$ & $0.88\pm{0.02}$ & $0.78\pm{0.03}$ & $0.84\pm{0.04}$ & $0.85\pm{0.02}$ & $0.67\pm{0.03}$ \\
0.054 & $0.88\pm{0.04}$ & $0.89\pm{0.01}$ & $0.80\pm{0.02}$ & $0.86\pm{0.04}$ & $0.86\pm{0.01}$ & $0.71\pm{0.03}$ \\
0.183 & $0.89\pm{0.04}$ & $0.92\pm{0.01}$ & $0.85\pm{0.02}$ & $0.87\pm{0.04}$ & $0.89\pm{0.01}$ & $0.79\pm{0.02}$ \\
0.209 & $0.94\pm{0.03}$ & $0.92\pm{0.01}$ & $0.83\pm{0.03}$ & $0.93\pm{0.03}$ & $0.90\pm{0.02}$ & $0.77\pm{0.03}$ \\
0.295 & $0.92\pm{0.03}$ & $0.92\pm{0.01}$ & $0.80\pm{0.03}$ & $0.90\pm{0.03}$ & $0.89\pm{0.01}$ & $0.71\pm{0.03}$ \\
0.314 & $0.94\pm{0.03}$ & $0.92\pm{0.02}$ & $0.80\pm{0.02}$ & $0.93\pm{0.03}$ & $0.90\pm{0.01}$ & $0.68\pm{0.03}$ \\
0.600 & $0.87\pm{0.04}$ & $0.88\pm{0.02}$ & $0.80\pm{0.02}$ & $0.85\pm{0.04}$ & $0.84\pm{0.02}$ & $0.73\pm{0.03}$ \\
0.620 & $0.90\pm{0.03}$ & $0.89\pm{0.01}$ & $0.79\pm{0.03}$ & $0.88\pm{0.04}$ & $0.87\pm{0.02}$ & $0.69\pm{0.03}$ \\
0.654 & $0.87\pm{0.04}$ & $0.88\pm{0.01}$ & $0.75\pm{0.03}$ & $0.85\pm{0.04}$ & $0.85\pm{0.01}$ & $0.63\pm{0.03}$ \\
0.892 & $0.79\pm{0.05}$ & $0.86\pm{0.02}$ & $0.83\pm{0.03}$ & $0.77\pm{0.05}$ & $0.81\pm{0.02}$ & $0.77\pm{0.03}$ \\
0.914 & $0.86\pm{0.04}$ & $0.86\pm{0.01}$ & $0.85\pm{0.02}$ & $0.84\pm{0.04}$ & $0.83\pm{0.02}$ & $0.80\pm{0.03}$ 
\enddata
\end{deluxetable*}

\subsection{Correlating $f_{spot}$ Variability with Stellar Rotation} \label{sec:ffactorallnights}

\begin{figure}[ht]
\includegraphics[width=8.5cm, height=10cm]{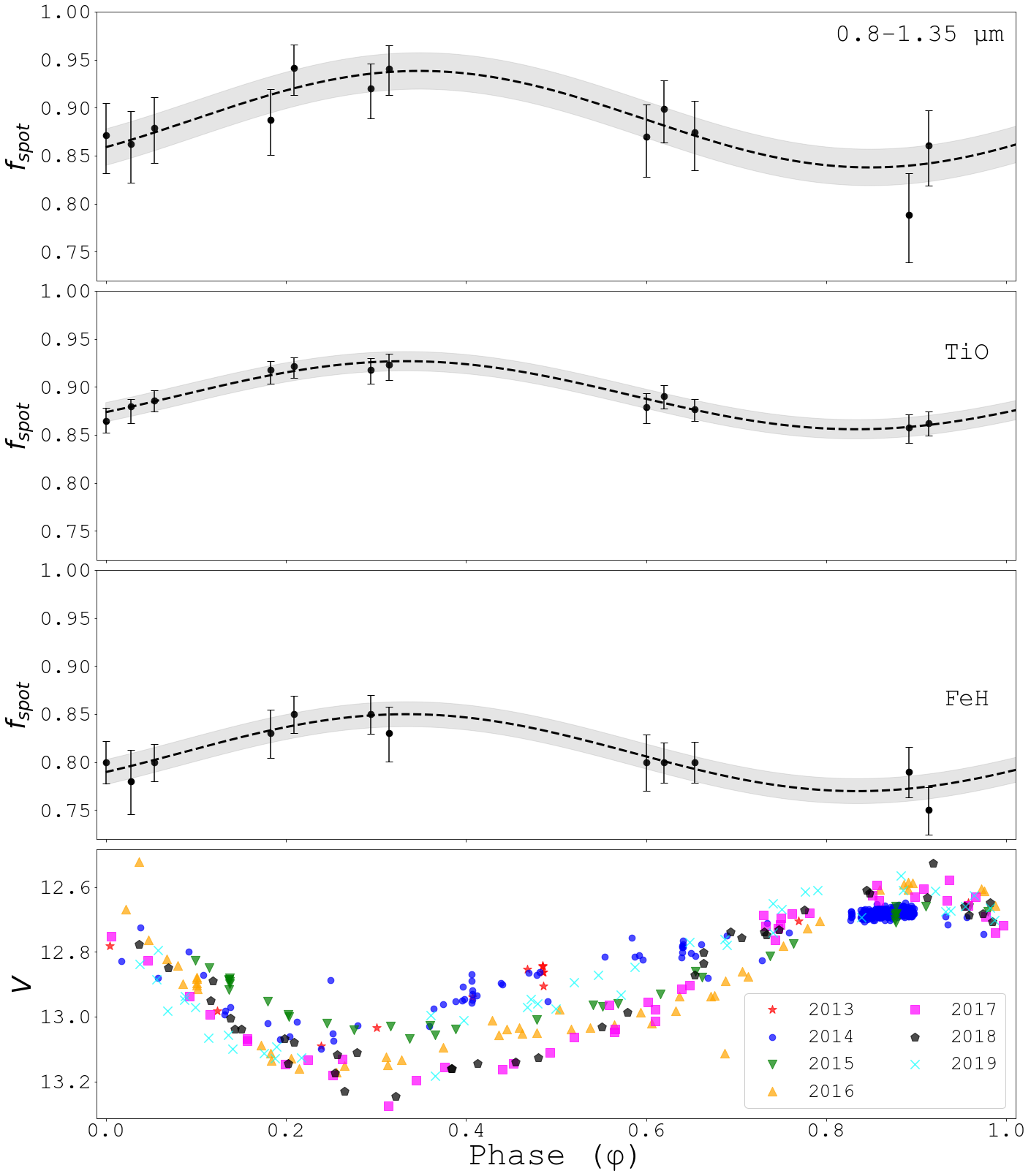}
\caption{(A) Plotted as a function of rotational phase are the spot filling factors from the best-fit model, K5V$_{p}$+M5V$_{s}$, to the $F_{0.8-1.35~\micron}$ indicator for all twelve LkCa~4 spectra with $A_V~=~0.3$. The dashed line represents a sinusoidal fit to the filling factors, while the gray shaded region represents the 1-$\sigma$ weighted uncertainty of the fit. (B) Same plot for TiO. (C) Same plot for FeH. (D) The phase-folded AAVSO $V$-band light curve made with seven years of data. \label{fig:fillingfactorplot}}
\end{figure}

In Figure~\ref{fig:fillingfactorplot}, we present nearly seven years of AAVSO $V$-band photometry collected between UT 2013 December 24 and UT 2020 October 17 along with the best-fit values for $f_{spot}$ for TiO and $F_{0.8-1.35\ \micron}$ as well as the corresponding values for FeH. All values have been phase-folded setting the time of our first observation taken at JD 2458488.71352 as $\phi$~=~0.0 and using $P_{rot}$~=~3.374~days \citep{Grankin2008}. The periodic variations in the spot filling factors are positively correlated with the periodic variability in the $V$-band light curve just as one would expect if the $V$-band variability were due to spots rotating with the surface of the star. Despite the differences in the absolute values of $f_{spot}$ derived from the TiO, $F_{0.8-1.35\ \micron}$, and FeH for models with similar values of $T_{phot}$ and $T_{spot}$, the variations within those values correlate with rotational phase (Figure~\ref{fig:fillingfactorplot}). 

Given what is known about spot lifetimes on young, active stars, we assume that the spot complex(es) on LkCa~4 are stable during the five nights over which these data were collected. Therefore, we conclude that the observed variability in the instantaneous $f_{spot}$ values is due to the rotation of the star and not due to periodic changes in the total spot coverage or the spot temperatures. In addition, we also rule out contributions to the variability from circumstellar material including phenomena such as inner disk warps and/or accretion flares \citep{Covey2021} based on the lack of evidence for a circumstellar disk and accretion activity in the LkCa~4 system \citep[e.g.,][]{Andrews2005}.

\subsubsection{Predicting $\Delta V$ from Best-fit Model Parameters}

Next we test whether or not our best-fit spotted star models can reproduce the $V$-band variability observed in the AAVSO light curve from UT 2019 given that the time frame aligns with the SpeX observations. We calculated the minimum to maximum $V$-band variability, $\Delta V$, using the following equation

\begin{equation}
\footnotesize
\Delta V_{calc}=\log\left(\frac{\int_{V}[(1-f_{max})F_{\lambda}(T_{p})+f_{max}F_{\lambda}(T_{s})]S_V(\lambda)d\lambda}{\int_{V}[(1-f_{min})F_{\lambda}(T_{p})+f_{min}F_{\lambda}(T_{s})]S_V(\lambda)d\lambda}\right)
\end{equation}

\noindent
where $F_{\lambda}$($T_p$) and $F_{\lambda}$($T_s$) are the flux densities of the photosphere and spot components, respectively, $f_{max}$ and $f_{min}$ are the maximum and minimum spot filling factors, and $S_V$($\lambda$) is the transmissivity of the Johnson-Cousins $V$-band filter as defined in the General Catalog of Photometric Data and revised by \citet{Mann2015}. 

We used BT-Settl(CIFIST) \citep{allard2014} synthetic spectra over the relevant wavelengths to represent the $F_{\lambda}$($T_p)$ and $F_{\lambda}$($T_s)$ contributions to the $V$-band magnitude. We calculated four values of $\Delta V_{\text{calc}}$, one for each of the $f_{spot}$ values derived from the $F_{0.8-1.35 \micron}$ and TiO indicators for models: K7V$_{p}$+M5V$_{s}$ and K5V$_{p}$+M5V$_{s}$. Values for the mean spot filling factor, $f^{mean}_{spot}$, and its amplitude, $f^{amp}_{spot}$, were extracted from a sinusoidal fit to the phase-folded $f_{spot}$ curves in Figure~\ref{fig:fillingfactorplot}. These values are listed in Table~\ref{tab:deltavpredict}. 

The observed value for the min-to-max variability in the 2019 AAVSO light curve, $\Delta {V_{\text{obs}}}~=~0.526\pm{0.032}$, agrees to within the uncertainties of $\Delta V_{\text{calc}}$~=~$0.56^{+0.10}_{-0.07}$ and $\Delta V_{\text{calc}}$~=~$0.46^{+0.09}_{-0.06}$ calculated for the $f_{spot}$ values associated with the TiO indicator and the best-fit models, K5V$_{p}$+M5V$_{s}$ and K7V$_{p}$+M5V$_{s}$, respectively. The $f^{amp}_{spot}$ values constrained by the $F_{0.8-1.35 \micron}$ indicator give a slightly larger $\Delta V_{\text{calc}}$ than the observed value for both models. The large $f^{mean}_{spot}$ values combined with a larger $f^{amp}_{spot}$ leads to a significant increase in the magnitude of the variation placing the corresponding $\Delta V_{\text{calc}}$ values slightly above the observed value.

\begin{deluxetable}{ccccc}[ht]
\setlength{\tabcolsep}{4.5pt}
\renewcommand{\arraystretch}{1.2}
\tabletypesize{\footnotesize}
\tablecaption{$f_{spot}^{\rm mean}$, $f_{spot}^{\rm amp}$, and $\Delta V_{\text{calc}}$ Values \label{tab:deltavpredict}}
\tablehead{
\colhead{Indicator} & \colhead{Model} &\colhead{$f_{spot}^{\rm mean}$} &\colhead{$f_{spot}^{\rm amp}$} & \colhead{$\Delta V_{\text{calc}}$} 
}
\decimals
\startdata
$F_{0.8-1.35\micron}$ & K5V$_{p}$+M5V$_{s}$   & $0.89\pm{0.02}$ & $0.05\pm{0.02}$ & $0.80^{+0.15}_{-0.11}$\\
TiO                   & \nodata             & $0.89\pm{0.01}$ & $0.04\pm{0.01}$ & $0.56^{+0.10}_{-0.07}$\\
\hline
$F_{0.8-1.35\micron}$ & K7V$_{p}$+M5V$_{s}$   & $0.87\pm{0.02}$ & $0.06\pm{0.02}$ & $0.66^{+0.13}_{-0.09}$\\
TiO                   & \nodata             & $0.86\pm{0.01}$ & $0.04\pm{0.02}$ & $0.46^{+0.09}_{-0.06}$\\
\enddata
\end{deluxetable}

The agreement between the amplitudes of variations in the spot sizes and the value of $\Delta {V}$ for the TiO indicator provides a reasonable consistency check for the two-temperature model parameters $f_{spot}$, $T_{spot}$, and $T_{phot}$.

\subsubsection{Year-to-year Variations of Spot Coverage}

The 2013-2019 AAVSO light curve presented in Figure~\ref{fig:fillingfactorplot}(d) possesses a few interesting aspects that suggest the modulation and possible evolution of the spot complexes during this time period. The phase-folded light curve possesses a vertical width of $\delta V$~=~0.1-0.2 mag at all phases, though it appears to be largest near $\phi$~=~0.5. Such a modulation in $\delta V$ indicates that the average spot coverage of the stellar surface has changed over this time frame. The slight asymmetry of the variations in the vertical width, particularly between $\phi$~=0.2 and 0.4 suggests an offset or shift in rotational phase, likely indicating a slight alteration in the timing of the minima and maxima over year-long timescales. We estimate the shift in rotational phase to vary between $\delta \phi \sim$~0.008 and $\delta \phi \sim$~0.084 over the 7yr period. Given the small uncertainties in the photometric measurements, these variations in the phased light curve are likely real features. We do not interpret these phase shifts necessarily as a change in the accepted 3.374 day rotation period of the star. Instead, we believe that they are most likely the result of the spots or spot complexes forming and dissipating at or migrating to different latitudes and/or longitudes. Similar and even more substantive secular changes to the spot coverage were highlighted in three decades of LkCa~4 observations (UT 1986-2016) presented in GS17.

\begin{deluxetable*}{ccccRcccc}
\setlength{\tabcolsep}{5pt}
\tabletypesize{\footnotesize}
\tablecaption{AAVSO $V$-band Photometry, $f^{\rm min}_{spot}$$\leftrightarrow$$f^{\rm max}_{spot}$, and $f_{phot}^{\rm max}$/$f_{phot}^{\rm min}$ Ratios  \label{tab:phottab}}
\tablehead{\colhead{UT Year} & \colhead{N$_{\rm obs}$\tablenotemark{a}} & \colhead{$V_{\rm mean}$} & \colhead{$\Delta$ $V_{\rm obs}$} & \colhead{Phase Shift\tablenotemark{b}} & \colhead{$f_{spot}^{\rm min}$$\leftrightarrow$$f_{spot}^{\rm max}$\tablenotemark{c}} & \colhead{$f_{phot}^{\rm max}$/$f_{phot}^{\rm min}$\tablenotemark{d}} & \colhead{$f_{spot}^{\rm min}$$\leftrightarrow$$f_{spot}^{\rm max}$\tablenotemark{c}} & \colhead{$f_{phot}^{\rm max}$/$f_{phot}^{\rm min}$\tablenotemark{d}}\\
\colhead{} & \colhead{} & \colhead{(mag)} & \colhead{(mag)} & \colhead{$\Delta \phi$} & \colhead{(K5V$_p$+M5V$_s$)}  & \colhead{(K5V$_p$+M5V$_s$)} & \colhead{(K7V$_p$+M5V$_s$)}& \colhead{(K7V$_p$+M5V$_s$)}
}
\decimals
\startdata
2013 & 14   & 12.825$\pm{0.015}$ & 0.406$\pm{0.053}$ & -0.039$\pm{0.014}$  & 0.87-0.92 & 1.63$^{+0.21}_{-0.13}$ & 0.84-0.91 & 1.78$^{+0.22}_{-0.14}$ \\
2014 & 330 & 12.835$\pm{0.003}$ & 0.325$\pm{0.007}$ &  0.008$\pm{0.004}$  & 0.87-0.91 & 1.44$^{+0.13}_{-0.08}$ & 0.85-0.90 & 1.50$^{+0.13}_{-0.08}$ \\
2015 & 36   & 12.880$\pm{0.005}$ & 0.381$\pm{0.013}$ &  0.076$\pm{0.005}$  & 0.86-0.91 & 1.56$^{+0.16}_{-0.10}$ & 0.83-0.90 & 1.70$^{+0.18}_{-0.12}$ \\
2016 & 54   & 12.923$\pm{0.016}$ & 0.463$\pm{0.048}$ &  0.084$\pm{0.014}$  & 0.85-0.91 & 1.56$^{+0.19}_{-0.12}$ & 0.80-0.90 & 2.00$^{+0.25}_{-0.09}$ \\
2017 & 45   & 12.932$\pm{0.006}$ & 0.608$\pm{0.018}$ &  0.053$\pm{0.004}$  & 0.83-0.92 & 2.13$^{+0.38}_{-0.23}$ & 0.79-0.91 & 2.33$^{+0.38}_{-0.24}$ \\
2018 & 40   & 12.916$\pm{0.007}$ & 0.593$\pm{0.018}$ &  0.051$\pm{0.005}$  & 0.84-0.92 & 2.00$^{+0.33}_{-0.20}$ & 0.80-0.91 & 2.22$^{+0.35}_{-0.22}$ \\
2019 & 43   & 12.879$\pm{0.010}$ & 0.526$\pm{0.032}$ &  0.000$\pm{0.008}$  & 0.84-0.94 & 2.67$^{-0.42}_{+0.83}$ & 0.81-0.92 & 2.38$^{-0.46}_{-0.28}$
\enddata
\tablenotetext{a}{Number of AAVSO observations made during the calendar year.}
\tablenotetext{b}{Shift in rotational phase measured with respect to 2019.}
\tablenotetext{c}{Range of $f_{spot}$ values calculated from $V$-band measurements.} 
\tablenotetext{d}{Photospheric filling factor ratios for the corresponding spot filling factors.}
\end{deluxetable*}

\begin{figure}
\centering
\includegraphics[width=8.5cm, height=10cm]{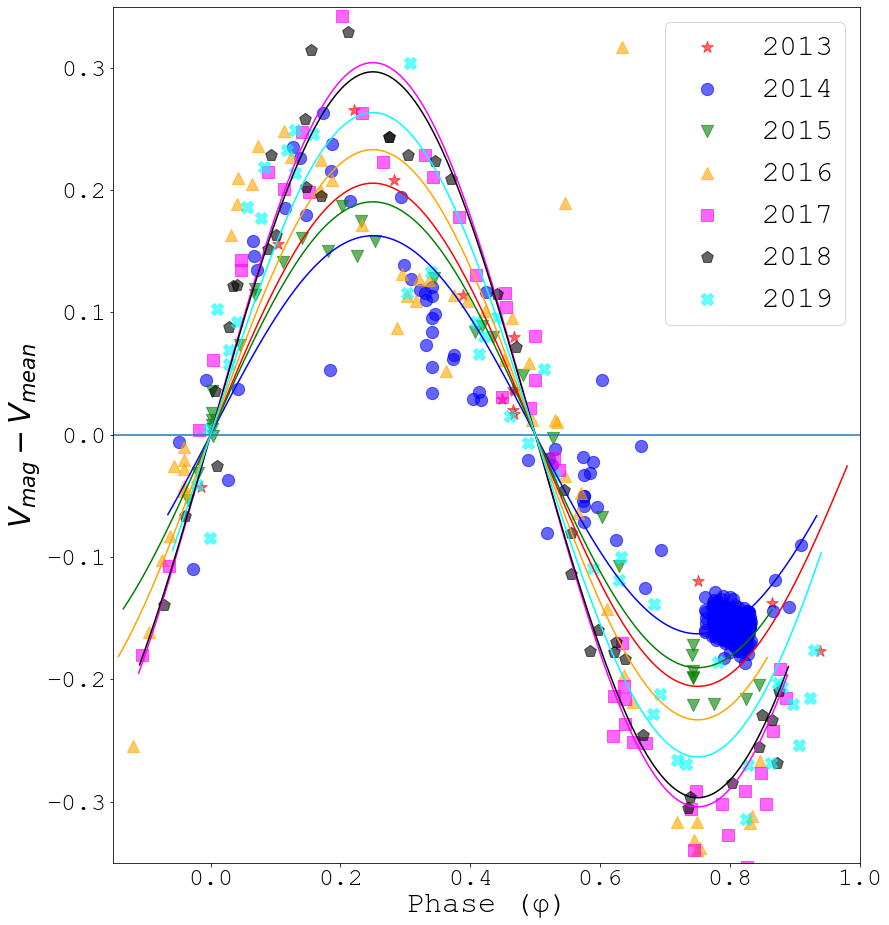}
\caption{Plotted are seven phase-folded AAVSO light curves from 2013-2019 representing each calendar year. The mean $V$-band magnitude for each year has been subtracted. Shifts in rotational phase have been removed. The solid lines represent the best-fit sinusoidal functions. The changes in the full photometric variability (minimum to maximum) range between 0.30 and 0.60 mag. \label{fig:yearlightucrves}}
\end{figure}

\begin{figure}
\includegraphics[width=8.5cm, height=10cm]{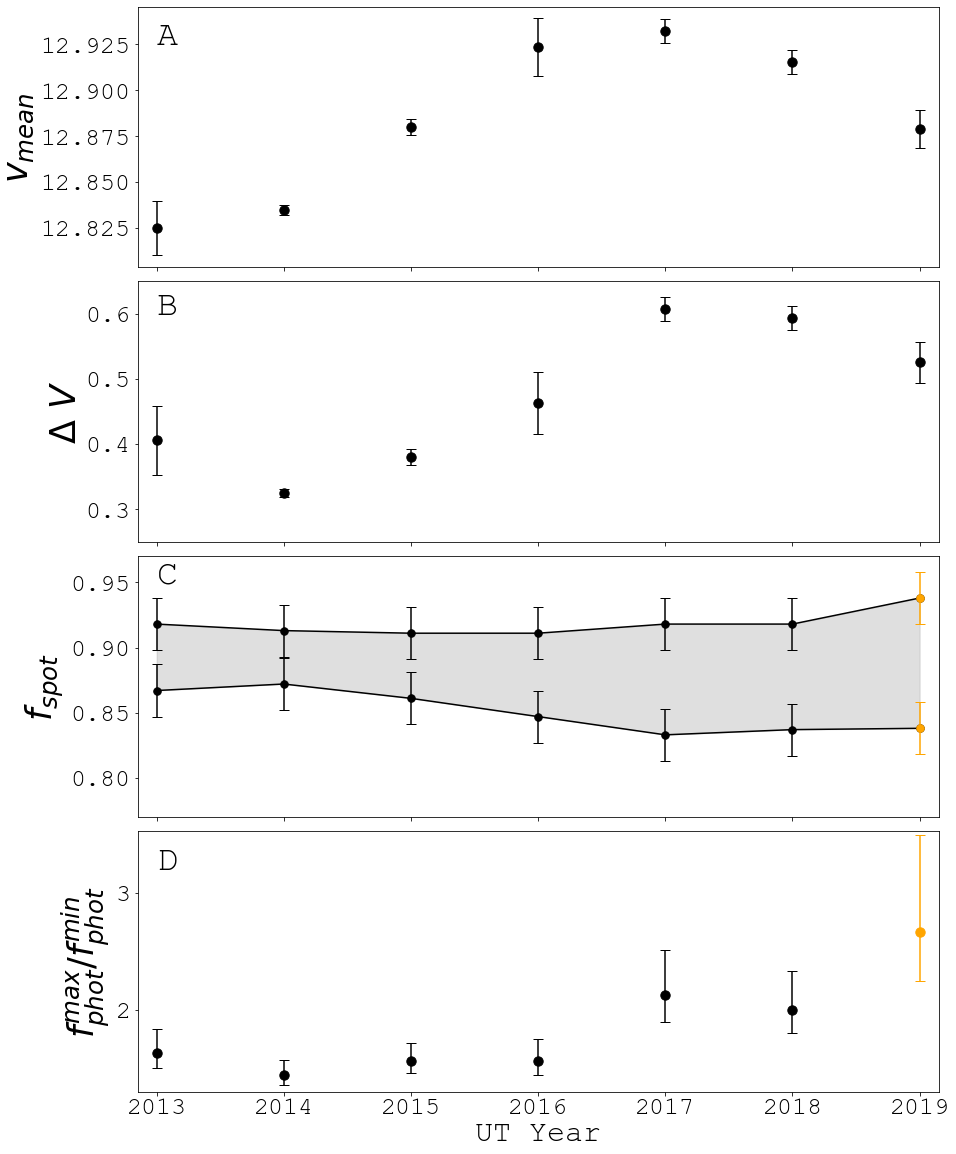}
\caption{(A) Mean visual magnitudes plotted as a function of year. (B) The min-to-max $V$-band variability plotted as a function of year. (C) The minimum and maximum spot filling factors for every year assuming a K5V$_{p}$+M5V$_{s}$ model. Shaded area represents the total parameter space for $f_{spot}$. The orange data points for 2019 were determined directly from model fits to spectroscopic data. (D) Ratios of maximum and minimum photospheric filling factors plotted for each year also for a K5V$_{p}$+M5V$_{s}$ model. Orange data points are same as in (C). (A-D) Black data points represent quantities measured directly from AAVSO photometry or a combination of spectroscopic models and photometric data. \label{fig:photfitplot}}
\end{figure}

Working from the photometry and assuming that the $T_{phot}$ and $T_{spot}$ remain constant over time, we can use $V_{\rm mean}$ and $\Delta V_{\rm obs}$ in the AAVSO data to estimate $f^{\rm mean}_{spot}$ and $f^{\rm amp}_{spot}$ in previous years. Assuming that the spot coverage is relatively constant on a year-long time scale, the AAVSO photometry from UT 2013-2019 was divided in 1yr periods. In Figure~\ref{fig:yearlightucrves}, we present each of the 7 yr long light curves with a best-fit sinusoidal function determined using the same minimization and MCMC algorithm adopted for the spectral fitting. For each light curve, $V_{\rm mean}$ has been subtracted to permit a direct comparison of the year-to-year variations of the amplitude. In addition, small shifts in rotational phase measured relative to the minima and maxima of the UT 2019 curve also have been removed. The values for $V_{\rm mean}$, $\Delta V_{\rm obs}$, and $\Delta \phi$ are listed in Table~\ref{tab:phottab}.

In order to determine $f^{\rm mean}_{spot}$ and $f^{\rm amp}_{spot}$ from previous years of $V$-band data, we associate the value for $f^{mean}_{spot}$ derived from our models with $V_{\rm mean}$ from 2019. Differences between the UT 2019 $V_{\rm mean}$ and previous year $V_{\rm mean}$ are used to determine the values of $f^{\rm mean}_{spot}$. The $\Delta V$ value measured from the light curve for a previous year can then be used to determine $f^{\rm amp}_{spot}$ (see Table~\ref{tab:phottab}).

Also presented are the ratios of the maximum to minimum \textit{photospheric} filling factors, $f_{phot}^{\rm max}$/$f_{phot}^{\rm min}$. While directly related to the ratio of the values of $f_{spot}$, we include this value to highlight the strong correlation between $\Delta V$ and the changes to $f_{phot}$. One can understand the nature of this correlation better by considering two stars that have the same $T_{phot}$ and $T_{spot}$, but possess the different mean filling factors of $f^{\rm mean}_{spot}~=~0.7$ and 0.8. For both stars, assume that the filling factors vary by the same amount, $\Delta f_{spot}$~=~0.1 or 10\%, such that the ranges of filling factors are 0.65~$\leq$~$f_{spot}$~$\leq$0.75 and 0.75~$\leq$~$f_{spot}$~$\leq$0.85, respectively. The star with the larger spot coverage would appear dimmer on average with a higher $V_{\rm mean}$ and would exhibit a larger amplitude of variability despite the same change in $f_{spot}$. The reason being that the photospheric emission dominates the brightness of the star in the $V$-band of even the most heavily spotted stars. Therefore, the amplitude of variability depends most directly on the ratio of the maximum and minimum filling factors of the photosphere, $f_{phot}^{\rm max}$/$f_{phot}^{\rm min}$ and less on the absolute change in $f_{spot}$. Hence, a 10\% change in $f_{spot}$ for the star with a larger $f^{\rm mean}_{spot}$ leads to a larger fractional change in the portion of the star covered by the hotter photospheric region. Such a change leads to a larger overall dimming/brightening of the star and a larger amplitude of variability.

The secular changes present in the AAVSO data are depicted in Figure~\ref{fig:photfitplot}, which plots the values of (a) $V_{\rm mean}$, (b) $\Delta V_{\rm obs}$, (c) the extrapolated minimum and maximum values for $f_{spot}$, and (d) $f_{phot}^{\rm max}$/$f_{phot}^{\rm min}$ by year. Within uncertainties, we find that the star is brightest when it displays the smallest $\Delta V$. These values for the $V_{\rm mean}$, the $\Delta V$ values, and the filling factor ratios agree with those of GS17, where a roughly twofold increase in the photospheric filling factor is required to cause a $\Delta V=0.6$.
 
\section{CO Spectral Index Temperatures} \label{CO}

The SpeX SXD spectra provide excellent wavelength coverage of the $K$ band, which includes several strong CO bandheads. The strength of the shortest wavelength CO band at 2.29~\micron\ is used as an indicator for $T_{eff}$, [Fe/H], and $\log~g$ and often applied to unresolved spectra of stellar populations in clusters and galaxies. \citet{marmolqueralto} updated the definition of a CO spectral index for the 2.29~$\micron$ feature, which they designate as $D_{\text{CO}}$. Given the impact of spots on spectral type determinations of heavily spotted PMS stars, we apply the $D_{\text{CO}}$ index to our spectra of LkCa~4 for comparison.

Following \citet{marmolqueralto}, we measure the $D_{\text{CO}}$ index as the ratio between the average fluxes in two spectral windows in the continuum near the feature
($\lambda_{cont1}$~=~2.2460-2.2550~$\micron$ and $\lambda_{cont2}$~=~2.2710-2.2770~$\micron$) and the average flux in the absorption band ($\lambda_{\text{CO}}$~=~2.288-2.3010~$
\micron$).
\vspace{0.0cm}

\begin{deluxetable}{ccc}
\setlength{\tabcolsep}{10pt}
\tablecaption{$D_{\text{CO}}$ Index Values \label{tab:dcotab}}
\tablehead{
\colhead{Phase}     & \colhead{$D_{\text{CO}}$} & \colhead{Temperature}\\
\colhead{($\phi$)}  &                           & \colhead{(K)}
}
\decimals
\startdata
0.000 & 1.1130$\pm{0.0079}$ & 3721$\pm{188}$\\
0.028 & 1.1089$\pm{0.0094}$ & 3826$\pm{233}$\\
0.054 & 1.1142$\pm{0.0081}$ & 3691$\pm{189}$\\
0.183 & 1.1071$\pm{0.0066}$ & 3874$\pm{171}$\\
0.209 & 1.1086$\pm{0.0097}$ & 3834$\pm{241}$\\
0.295 & 1.1087$\pm{0.0071}$ & 3831$\pm{179}$\\
0.314 & 1.1074$\pm{0.0067}$ & 3866$\pm{173}$\\
0.600 & 1.1101$\pm{0.0075}$ & 3795$\pm{186}$\\
0.620 & 1.1115$\pm{0.0076}$ & 3759$\pm{185}$\\
0.654 & 1.1145$\pm{0.0054}$ & 3684$\pm{123}$\\
0.892 & 1.1147$\pm{0.0101}$ & 3679$\pm{231}$\\
0.914 & 1.1120$\pm{0.0065}$ & 3746$\pm{158}$
\enddata
\end{deluxetable}
\vspace{0.0cm}

\begin{figure}
\centering
\includegraphics[width=8cm, height=10cm]{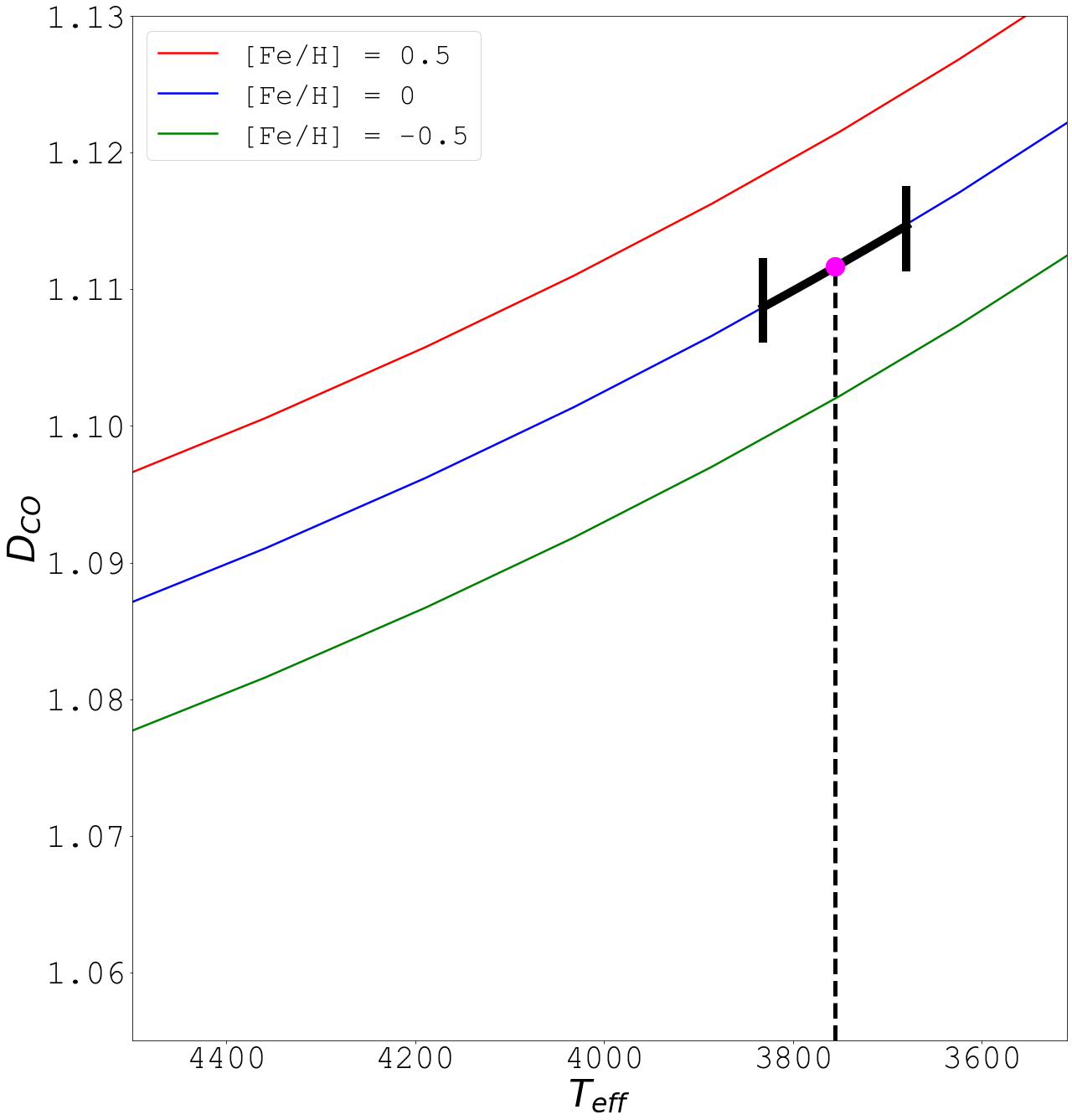}
\caption{$D_{\text{CO}}$ index fitting functions for dwarf stars with metallicities [Fe/H]~=~-0.5, 0.0, and 0.5 for 
$\log g~=~3.8$ are plotted as a function of temperature from \label{fig:dcoindex} \citet{marmolqueralto}. The 
filled circle (magenta) represents the weighted mean value for the $D_{CO}$ index. The error bar represents the 
uncertainty on the weighted mean.}
\end{figure}
\vspace{0.0cm}

$D_{\text{CO}}$ index values and uncertainties were calculated for each of the 12 observations of LkCa~4. The 
index values fall within the range of $1.1074-1.1147$ with a value for the weighted mean of 1.1111$\pm$0.0022. 
Using the empirical fitting functions provided by \citet{marmolqueralto}, we assume a metallicity of [Fe/H] = 0.0 
and $\log$~$g$~=~3.8 to find a corresponding value of $T_{eff}$~=~$3755\pm{76}$~K (see Figure~\ref{fig:dcoindex}). 
Table~\ref{tab:dcotab} lists all of the $D_{\text{CO}}$ and corresponding $T_{eff}$ values. 

The mean temperature is two subclasses later than a K7V (4100~K), the spectral type reported in \citet{Donati2014}. 
The value more closely matches the temperature of $3670~K$ determined by \citet{herczeg2014}. The $D_{\text{CO}}$ 
temperature, lying between the photospheric and spot temperatures and in relatively good agreement with 
\citet{herczeg2014} value, suggests that the $D_{\text{CO}}$ index is somewhat sensitive to the presence of the 
cool spots. However, we find in the following analysis of spot-corrected temperatures that the $D_{\text{CO}}$ 
temperature is significantly warmer.

\section{Spot-corrected $T_{eff}$, $L_\star$, and SPOTS Evolutionary Models} \label{sec:spotsmodels}

Next we calculate spot-corrected values for $T_{eff}$ and $L_\star$ from the values derived for $f^{\rm 
mean}_{spot}$, $T_{phot}$, and $T_{spot}$. First, we adopt $f^{\rm mean}_{spot}$ values and uncertainties from the 
best-fit composite models, K7V$_p$+M5V$_s$ and K5V$_p$+M5V$_s$, as estimates of the \textit{total} spot coverage of 
LkCa~4 with the understanding that a portion of the star is never visible due to the inclination of the stellar 
rotation axis ($i~=~35\degree$; GS17). Values for $T_{eff}$ were calculated using the following

\begin{equation}
    T_{eff} = \left[T_{phot}^4(1-f_{spot})+T^4_{spot}f_{spot}\right]^{0.25}
\end{equation} \label{eqn:teff}

\noindent
for the model parameters associated with spot indicators: $F_{0.8-1.35\ \micron}$ and TiO. Adopting a stellar 
radius of 2.3~$R_\odot$ (GS17), we calculated the corrected values for $L_\star$. The uncertainties on $T_{eff}$ 
were estimated by assuming one-half of a subclass uncertainty on the spectral types and associated temperatures for 
both parameters $T_{spot}$ and $T_{phot}$. The resulting extrema of $T_{eff}$ values were then used to calculate 
the upper and lower bounds for the corrected $L_\star$ values. The values derived for $T_{eff}$ and $L_\star$ are 
listed in Table~\ref{tab:HRdata}.

\begin{figure*}
\centering
\includegraphics[width=14.0cm, height=10cm]{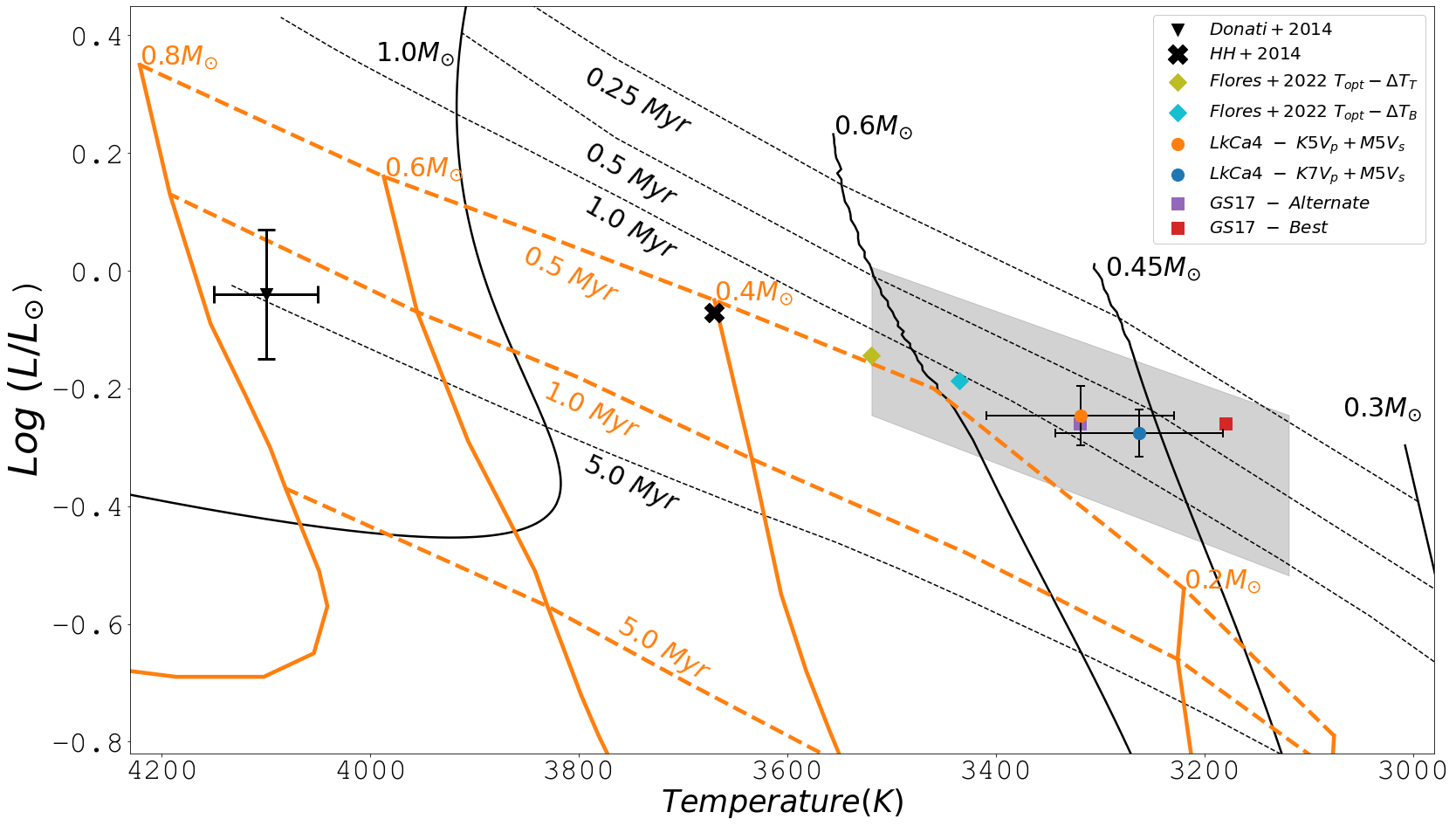}
\caption{Plotted are the spot-corrected $T_{eff}$ and L$_\star$ values associated with the K5V$_p$+M5V$_s$ (orange 
circle) and K7V$_p$+M5V$_s$ (blue circle). Error bars on the values come from assuming a half-subclass uncertainty 
in $T_{phot}$ and $T_{spot}$ and from the uncertainty in the mean $f_{spot}$. Overlaid are isochrones (dashed) 
and evolutionary tracks (solid) from \citet{Baraffe2015} (orange) and \citet{Somers2020} (black) with 
$f_{spot}~=~0.85$. The gray parallelogram encloses the full range of $T_{eff}$ and $L_\star$ values from the 
\textit{instantaneous} filling factors with uncertainties. The solitary black triangle represents the values from 
\citet{Donati2014}. The black cross at $T=3670\ K$ is from \citet{herczeg2014}. The green ($\Delta T_{opt-ir}$) 
and cyan ($\Delta T_B$) diamonds represent the $T_{eff}$ values derived from Flores+2022 empirical relations. The 
red and purple squares are the ``best" ($T_{phot}=4100\ K$, $T_{spot}=2750\ K$, $f_{spot}=0.80$) and ``alternate" 
($T_{phot}=3100\ K$, $T_{spot}=3000\ K$, $f_{spot}=0.80$) estimates from GS17. \label{fig:trackscomparisonffactor}}
\end{figure*}

In Figure~\ref{fig:trackscomparisonffactor}, the corrected $T_{eff}$ and $L_\star$ values with the uncertainties 
obtained from the $F_{0.8-1.35\ \micron}$ indicator are placed on an H-R diagram alongside literature values taken 
from \citet{Donati2014}, \citet{herczeg2014}, GS17, and two values derived using empirical relationships (detailed 
below) found in \citet{Flores2022}. While our two data points with uncertainties reflect only the mean value of 
$f_{spot}$ with its uncertainties, the gray parallelogram encompasses the entire parameter space associated with 
$f^{\rm min}_{spot}$ and $f^{\rm max}_{spot}$ measured over the full rotation of the star. Therefore, any single-epoch observation used to correct $T_{eff}$ and $L_\star$ should place the star within the gray parallelogram.

Overlaid are the isochrones and tracks from the \citet{Baraffe2015} standard evolutionary models (orange), as well 
as those from the Stellar Parameters of Tracks with Starspots (SPOTS) models for $f_{spot}$~=~0.85 \citep[black,][]
{Somers2020}. The SPOTS evolutionary models incorporate the structural effects of large cool starspots by 
accounting for phenomena such as the inhibition of convection by strong magnetic fields and the impact spots have 
on the pressure of the stellar photospheres. SPOTS evolutionary tracks are calculated for low-mass stars with spot 
filling factors as large as 0.85, making LkCa~4 a perfect candidate for comparison to these models.

First, we note the position of the non-corrected optical $T_{eff}$ and $L_\star$ values from Donati et al.\ in 
Figure~\ref{fig:trackscomparisonffactor}, which places LkCa~4 between the 0.7 and 0.8~$M_\odot$ standard 
evolutionary tracks with an age in the range of $\sim$1-3~Myr. The cooler value for $T_{eff}$ from 
\citet{herczeg2014} shifts the masses down to 0.4~$M_\odot$ and the age to $\sim$~0.5~Myr. The rest of the values 
plotted are in some sense corrected for the existence of spots or cooler regions of the star and shift the values 
of $T_{eff}$ farther to the right and slightly downward on the H-R diagram.

Relative to the Baraffe+2015 standard evolutionary models, the corrected values yield lower mass ranges of 0.25-
0.30~$M_{\odot}$ with ages less than 0.5~Myr. These new values represent a shift in mass by a factor of two or 
slightly greater and a decrease in age by a factor between two and six. 

Comparing the two sets of evolutionary models, the tracks for the SPOTS models are shifted to cooler temperatures 
and higher luminosities. Given that the corrected $T_{eff}$ and $L_\star$ moves the star down and to the right, it 
appears that accounting for the evolutionary effects of spots mitigates some of the shift toward lower masses and 
younger ages. As such, the corrected placement of LkCa~4 in relation to the $f_{spot}$~=~0.85 SPOTS models 
increases the mass to a range of 0.45-0.60~$M_\odot$ and the age to a range of 0.5-1.25~Myr, relative to the 
Baraffe+2015 models.

\begin{figure}
\includegraphics[width=8.5cm, height=10cm]{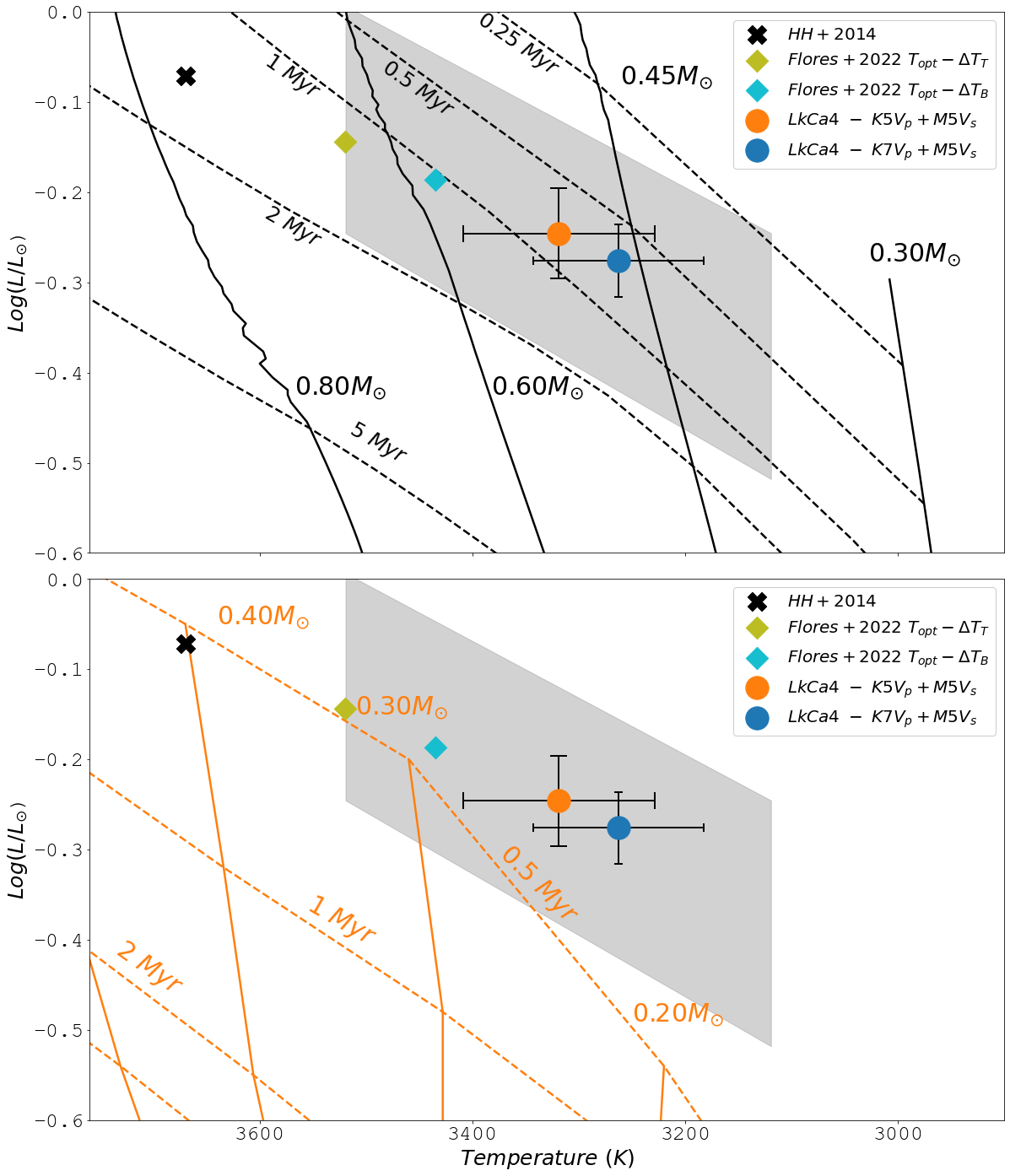}
\caption{(Top) A close-up view of the corrected $T_{eff}$ and L$_\star$ values for LkCa~4 with respect to the 
SPOTS models with $f_{spot}$~=~0.85 (black lines). The filled orange circle corresponds to the K5V$_p$+M5V$_s$ 
model. The filled blue circle corresponds to the K7V$_p$+M5V$_s$ model. (Bottom) Same comparison for the Baraffe+2015 
models (orange lines). In both plots, the gray parallelogram shows the parameter space for spot-corrected values 
over the full range of instantaneous filling factors. Plotted values are the same as in 
Figure~\ref{fig:trackscomparisonffactor} with the exclusion of the Donati et al.\ data point. 
\label{fig:hrdiagram}}
\end{figure}

We now turn our attention to the two data points obtained using the empirical relationships derived in 
\citet{Flores2022}. Based on the correlations these authors find between the optical and infrared temperatures as 
well as the optical temperatures and the strengths of the stellar magnetic fields, they develop two empirical 
equations for the shifts in temperature: (1) $\Delta T_{opt-ir}~=~0.36T_{opt} - 1170~K$ and (2) $\Delta T_{B}~=~206B 
- 135~K$. Note that the uncertainties in the Flores+2022 relations are large, so we chose only to use the median 
values. For both equations, the values for $\Delta T_{opt-ir}$ and $\Delta T_{B}$ are to be subtracted from 
$T_{opt}$ to determine $T_{eff}$. Following Flores et al., we adopted $T_{opt}$~=~3670~K for LkCa~4 from 
\citet{herczeg2014}. For the first relationship, we find a value of $T_{eff}~=~3520$~K and correspnding to $Log\ 
(L/L_\odot)=-0.144$. For the second relationship, we assumed the median $B$-field strength of 1.8~kG reported by 
Flores et al., likely to be a conservative value for LkCa~4, and find $T_{eff}$~=~3435~K with $Log\ (L/L_\odot)=- 
0.187$. As in all other cases, the associated values of $L_\star$ were calculated using $R_\star$~=~2.3~$R_\odot$. 
Given the large uncertainties on the empirical relations, we are somewhat surprised to see such reasonable 
agreement between the values these relations predict and the parameter space we find using our two-temperature 
models. Figure~\ref{fig:hrdiagram} gives a closer view of the corrected placement of LkCa~4, the GS17 values, and 
those derived from the Flores+22 empirical relations with respect to both SPOTS and standard evolutionary models.

For a subset of stars in the \citet{Flores2022} sample, dynamical masses were available from Atacama Large Millimeter/submillimeter Array observations and 
were compared to the masses inferred from both optical and infrared temperatures using the \citet{Feiden2016} 
evolutionary models which incorporate magnetic effects. They find that the masses inferred from infrared 
temperatures are much closer to the dynamical masses of the stars than masses inferred from the optical 
temperatures. However, for the lowest mass stars in their sample, 0.25-0.4~$M_\odot$, the infrared masses are also 
overestimations when compared to the dynamical masses by as much as 50\%. The interesting and potentially paradigm-shifting implications of this work is that the ``spot" component, $F_\lambda$($T_{spot}$), of the two-temperature 
models appears to be a better approximation of the star's spectral type than the photospheric component, 
$F_\lambda$($T_{phot}$). Perhaps, this shift in thinking is unsurprising for a PMS star such as LkCa~4 in which 
the ``spot" temperatures were shown to dominate over two-thirds of the stellar surface. The warmer component, in 
these cases being confined to comparatively smaller regions of the star, behaves more as a \textit{warm} spot on a cooler 
star than as the stellar photosphere. 

\vspace{0.0cm}

\begin{deluxetable}{ccccc}
\setlength{\tabcolsep}{5pt}
\tablecaption{Spot-corrected $T_{eff}$ and $L_\star$ Values \label{tab:HRdata}}
\tablehead{
\colhead{Indicator} & \colhead{Model} & \colhead{$T_{eff}(K)$} & \colhead{$Log\ (L/L_\odot)$}
}
\decimals
\startdata
$F_{0.8-1.35\micron}$   & K5V$_p$+M5V$_s$ & $3319\pm{90}$ & $-0.25\pm{0.05}$\\
TiO                     & \nodata         & $3312\pm{80}$ & $-0.25\pm{0.04}$\\
$F_{0.8-1.35\micron}$   & K7V$_p$+M5V$_s$ & $3263\pm{80}$ & $-0.28\pm{0.04}$\\
TiO                     & \nodata         & $3272\pm{80}$ & $-0.27\pm{0.04}$\\
\enddata
\end{deluxetable}

The results of these studies strongly suggest that previous determinations of temperatures and luminosities for 
many young stars are inherently flawed since the optical spectra appear to be dominated by small, warm, non-representative regions of the stellar surface. For such stars, the masses and ages inferred from the warmer 
temperatures will represent overestimates of masses and ages alike, whether compared to standard evolutionary 
models or those that take into account magnetic effects and spots. Whether or not these results hint at a new 
physical phenomenon associated with these stars is unclear, but for sources with large discrepancies between 
optical and infrared colors and temperatures, the spread in masses and ages for objects in a given cluster will be 
quite large if left uncorrected.

\subsection{Impact of Short- and Long-term Variability on Mass and Age Estimates}

The short-term, periodic variability of the filling factors due to stellar rotation and the long-term variability 
associated with changes in the total spot coverage both impact the placement of a star on the H-R diagram and the 
SPOTS models chosen for comparison. The parameter space enclosed within the solid parallelograms in both 
Figures~\ref{fig:trackscomparisonffactor} and \ref{fig:hrdiagram} accounts for the full range of variability of the 
instantaneous filling factors observed over one full rotation of the star. The min-to-max amplitude of the short-term variability in the filling factors is $\sim$0.1. In the long-term $V$-band variability inferred from the 
AAVSO data (Figure~\ref{fig:yearlightucrves}) spanning nearly a decade and several thousand stellar rotations, we 
find a comparable level of variability in the total spot filling factor. The parameter space bounded by the 
parallelogram also constrains the range of ages and masses caused by long-term variations in spot coverage. 
Therefore, while the application of the SPOTS models is affected by changes to the spot coverage on the order of 
years, LkCa~4 suggests that over the period of the last decade these models would constrain the mass uncertainties 
to be $\Delta M_\star$ = ${0.10}$~$M_\odot$ and age uncertainties to be $\Delta \text{Age}$ = ${0.625}$~Myr.

\section{Summary \& Conclusion} \label{sec:Conclusion}
 
We illustrate the utility of multi-epoch, medium-resolution spectroscopy combined with two-temperature empirical 
composite models to accurately constrain (1) the photospheric and spot temperatures, (2) the visual extinction, and 
(3) the instantaneous and total spot filling factors for the heavily-spotted young star LkCa~4. Relying 
predominantly on the 0.8-1.35~\micron\ region of the SpeX spectra that is sensitive to interstellar extinction and 
possesses absorption bands associated with TiO and FeH, we fit spectral models of spotted stars with four 
different photospheric temperatures, eight spot temperatures, and 11 visual extinctions to each of the 12 
observations of LkCa~4 collected over five consecutive nights allowing $f_{spot}$ to vary freely between 0.0 and 
1.0. Minimizing $\chi^2$ over all possible models, we find two best-fit composite spectra, K5V$_p$+M5V$_s$ and 
K7V$_p$+M5V$_s$, and $A_V$~=~0.3 consistently provide the best fits. Night-to-night variations in the filling 
factors positively correlate with the historic AAVSO light curves for the system with a rotational period of 3.374 
days. Such a correlation demonstrates how multi-epoch spectroscopic observations with moderate resolution can 
detect the rotation of a spotted star and better constrain the total spot coverage than a single observation.

In addition, we have used the stellar parameters, $T_{phot}$, $T_{spot}$, and $f_{spot}$ returned by the best-fit 
composite models to predict the magnitudes of $V$-band variability observed in a similar time frame to when the 
spectroscopic data was collected and found good agreement. Whether we think of the star as possessing the warmer 
photosphere with a significant fraction of its surface mottled with cooler regions of suppressed convection 
activity or being a cooler star with warm spots, the two-temperature models simultaneously explain the observed 
spectroscopic and photometric variability of the source. Regardless, the observed correlation between magnetic 
field strengths, anomalous colors, and optical and infrared spectral type mismatches hint at a magnetic origin to 
these phenomena \citep{Flores2022}.  

Assuming that $F_\lambda$(${T_p}$) and $F_\lambda$(${T_s}$) remain fairly constant over year to decade-long 
timescales, the 7yr time period of AAVSO data studied indicates that the total spot filling factor does not 
change by more than 5-10\% over this time frame. Small shifts in the rotational phase over year-long time 
intervals indicate possible migration and evolution of the spots on this time frame.

In comparing the placement of the optical temperatures and luminosities and the corrected values to standard 
evolutionary models on the H-R diagram, we infer significantly lower masses and younger ages for LkCa~4; a result 
that agrees well with those from GS17. When compared to the SPOTS models, we find some of the shift to lower 
masses and younger ages is mitigated. However, the shifts are still significant and important in light of the 
spread in ages frequently observed for star forming regions. The range of the corrected values of $T_{eff}$ and 
$L_\star$ associated with single-epoch observations alone shows considerable spread in mass, $\Delta 
$M$_\star$~=~0.2~M$_\odot$ and $\Delta$~Age~$\approx$~1~Myr and makes the case for more multi-epoch studies of 
young spotted stars.

Given the apparent ubiquity of optical vs.\ infrared color and spectral type discrepancies for young, low-mass 
stars, characterizing the impact on previous measurements of $T_{eff}$ and $L_\star$ will be useful for revising 
and, possibly, refining age and mass inferences for most young stellar clusters. In general, these results should 
also problematize the notion that a single spectral type can be assigned to most young stars. For instance, the 
$D_{\text{CO}}$ index points to a $T_{eff}$ that more closely aligns with that determined from optical and NIR 
TiO features. This value is a few hundred Kelvin warmer than the corrected values from the two-temperature 
empirical composite models, but several hundred Kelvin cooler than the a spectral type based on both optical 
spectra and color indexes. 

The work of \citet{Flores2022} shows that masses inferred from infrared temperatures much more closely align with 
the dynamical masses of the star, yet they too are likely overestimates for the lowest mass stars in their sample. 
Their work highlights the need for more comparisons between mass inferences involving evolutionary models and 
dynamically determined stellar masses. Nonetheless, these results intriguingly point toward a fundamental shift 
in the phenomenological understanding of the color and spectral type discrepancies.

A collective effort by the star forming community combining multiwavelength and multi-epoch observations of 
sources in several nearby star forming regions would help to characterize total spot coverage, spot and 
photospheric temperatures, and determine spot-corrected values for $T_{eff}$ and $L_\star$ across the optical and 
infrared. Such studies would also permit the application of the newer SPOTS evolutionary models to an extensive 
sample of PMS stars. As illustrated by \citet{Flores2022}, important to these efforts will be the availability of 
dynamical masses, which will provide useful and necessary constraints on the models as well as additional 
confirmation or refutation of the phenomenon of large cool starspots. The availability of medium-resolution NIR spectrographs on 3 and 4 meter class telescopes makes this methodology an attractive approach to determining 
spot characteristics for a large number of nearby PMS.

The authors wish to thank Kevin Covey and Christian Flores for early and insightful discussions related to this 
work. We also thank spot expert Steve Saar for his insights related to the nature of spots and the magnetic 
effects on the spot indicators. We thank the staff at NASA's IRTF for generously supporting the observations 
presented herein. We thank the anonymous referee for helpful feedback that has certainly improved this work. We 
acknowledge with thanks the variable star observations from the AAVSO International Database contributed by 
observers worldwide and used in this research. H.M. acknowledges support from the National Science Foundation 
through the Keck Northeast Astronomy Consortium's REU program through grant AST-1950797. The authors wish to 
recognize and acknowledge the very significant cultural role and reverence that the summit of Maunakea has always 
had within the indigenous Hawaiian community. We are most fortunate to have the opportunity to conduct 
observations from this mountain. The authors also acknowledge a publication grant from Colgate Univerisity's Research Council.

\section{Software and third party data repository citations} \label{sec:cite}
\facilities{IRTF (SpeX)}


\software{astropy \citep{astropy2013}, SpexTool \citep{cushing2004}, matplotlib \citep{Hunter2007}, SciPy 
\citep{jones2014}, NumPy \citep{VanderWalt2011}, lmfit \citep{Newville2014}, emcee \citep{Foreman-Mackey2013}}
\vspace{1cm}

\restartappendixnumbering
\appendix
\section{Selection of Best-fit Model Parameters and Associated $\chi^2_{red}$ Values}

In order to determine the best-fit models and values for $F_\lambda$($T_{phot}$), $F_\lambda$($T_{spot}$), 
$f_{spot}$, and $A_V$, we probed a large parameter space spanning four photospheric templates, eight spot 
templates, 11 visual extinctions, and filling factors freely varying between 0.0 and 1.0. In 
Tables~\ref{tab:wholefittingK5} \& \ref{tab:wholefittingK7}, we report the best-fit $f_{spot}$ and associated 
$\chi^2_{\rm red}$ values determined from the three spot indicators for each of the 12 observations. 
Values are listed for the two best photospheric templates K5V$_p$ and K7V$_p$ combined with six of the eight spot 
templates. All have been calculated using $A_V$~=~0.3.

For both the $F_{\lambda}$(K5V$_{p}$) and $F_{\lambda}$(K7V$_{p}$) model fits, we find that the warmest (M1V and M2V) 
and coolest (M7V and M8V) spot templates produce the largest $\chi^2_{red}$ values and the 
poorest fits for both TiO and $F_{0.8-1.35\micron}$ spot indicators. Models constructed with M4V$_{s}$, 
M5V$_{s}$, and M6V$_{s}$ spot templates produce substantially better fits. For each of the six possible composite 
models and each of the three spot indicators, we calculated the mean $\chi^2_{red}$ values over the 12 
observations. The K5V$_{p}$+M5V$_{s}$ and K7V$_{p}$+M5V$_{s}$ models possess both the lowest mean $\chi^2_{\rm 
red}$ values (with the value for the K5V$_p$ models being marginally better than the K7V$_p$) and the smallest 
standard deviations from the mean suggesting these models are consistently the best fits to the 12 
observations spread over five nights.

As discussed in \ref{sec:fitting}, the FeH indicator does not easily distinguish between the best-fit 
models, selecting models with warm spots and extremely large filling factors (i.e., essentially single-temperature 
fits) and those with cooler spots and smaller filling factors equally over nearly all the epochs. In addition, 
the poorer fits made to the models using an M4V standard to model the spots seem to be an outlier and may reveal 
something about the nature of the standard (see Figure~\ref{fig:fehfits}).

\begin{splitdeluxetable*}{c|cr|cc|cc|cc|cc|cc|cc|cc|ccBcc|cc|cc|cc|cc|cc|cc|cc|cc}
\tabletypesize{\footnotesize}
\tablecaption{$f_{spot}$ \& $\chi^2_{\rm red}$ Values for K5V$_p$ Template Models\label{tab:wholefittingK5}}
\tablehead{
\colhead{Phase} & \multicolumn{6}{c}{K5V$_{p}$+M3V$_{s}$} & \multicolumn{6}{c}{K5V$_{p}$+M4V$_{s}$} & 
\multicolumn{6}{c}{K5V$_{p}$+M5V$_{s}$} & \multicolumn{6}{c}{K5V$_{p}$+M6V$_{s}$} &
\multicolumn{6}{c}{K5V$_{p}$+M7V$_{s}$} &
\multicolumn{6}{c}{K5V$_{p}$+M8V$_{s}$}\\
\colhead{}         & \twocolhead{$F_{0.8-1.35\ \micron}$} & \twocolhead{TiO}   & \twocolhead{FeH} & 
\twocolhead{$F_{0.8-1.35\ \micron}$} & \twocolhead{TiO}   & \twocolhead{FeH} & \twocolhead{$F_{0.8-1.35\ \micron}$}
& \twocolhead{TiO}   &  \twocolhead{FeH} & \twocolhead{$F_{0.8-1.35\ \micron}$} & \twocolhead{TiO}   &  
\twocolhead{FeH} & \twocolhead{$F_{0.8-1.35\ \micron}$} & \twocolhead{TiO}   & \twocolhead{FeH} & 
\twocolhead{$F_{0.8-1.35\ \micron}$} & \twocolhead{TiO} & \twocolhead{FeH} \\
\hline
\colhead{$\phi$}   & \colhead{$f_{spot}^{best}$}   & \colhead{$\chi^2_{\rm red}$}  & \colhead{$f_{spot}^{best}$}  
& \colhead{$\chi^2_{\rm red}$} & \colhead{$f_{spot}^{best}$}   & \colhead{$\chi^2_{\rm red}$}            & 
\colhead{$f_{spot}^{best}$}   & \colhead{$\chi^2_{\rm red}$} & \colhead{$f_{spot}^{best}$}   & 
\colhead{$\chi^2_{\rm red}$}            & \colhead{$f_{spot}^{best}$}   & \colhead{$\chi^2_{\rm red}$} & 
\colhead{$f_{spot}^{best}$} & \colhead{$\chi^2_{\rm red}$}   & \colhead{$f_{spot}^{best}$}   & 
\colhead{$\chi^2_{\rm red}$}   & \colhead{$f_{spot}^{best}$} & \colhead{$\chi^2_{\rm red}$} & 
\colhead{$f_{spot}^{best}$}   & \colhead{$\chi^2_{\rm red}$}            & \colhead{$f_{spot}^{best}$}   & 
\colhead{$\chi^2_{\rm red}$} & \colhead{$f_{spot}^{best}$} & \colhead{$\chi^2_{\rm red}$}   & 
\colhead{$f_{spot}^{best}$}   & \colhead{$\chi^2_{\rm red}$}   & \colhead{$f_{spot}^{best}$} & 
\colhead{$\chi^2_{\rm red}$} & \colhead{$f_{spot}^{best}$}   & \colhead{$\chi^2_{\rm red}$}            & 
\colhead{$f_{spot}^{best}$}   & \colhead{$\chi^2_{\rm red}$} & \colhead{$f_{spot}^{best}$} & \colhead{$\chi^2_{\rm 
red}$}   & \colhead{$f_{spot}^{best}$} & \colhead{$\chi^2_{\rm red}$}   
}
\decimals
\startdata
0.000 & 1.00 & 4.37 & 1.00 & 1.40 & 1.00 & 1.56 & 0.97 & 3.40 & 0.92 & 1.88 & 1.00 & 2.04 & 0.88 & 2.30 & 0.86 & 
1.50 & 0.80 & 1.65 & 0.84 & 2.74 & 0.88 & 2.34 & 0.76 & 1.41 & 0.86 & 3.48 & 0.87 & 2.59 & 0.73 & 2.14 & 0.83 & 
5.23 & 0.91 & 2.83 & 0.71 & 1.99 \\
0.028 & 1.00 & 3.65 & 1.00 & 1.43 & 1.00 & 1.17 & 0.96 & 3.35 & 0.93 & 1.58 & 1.00 & 1.53 & 0.87 & 2.21 & 0.87 & 
1.37 & 0.78 & 1.28 & 0.83 & 2.66 & 0.88 & 2.32 & 0.75 & 1.23 & 0.85 & 3.48 & 0.88 & 2.72 & 0.71 & 1.92 & 0.83 & 
4.78 & 0.92 & 2.76 & 0.69 & 1.87 \\ 
0.054 & 1.00 & 4.89 & 1.00 & 1.75 & 1.00 & 1.49 & 0.98 & 2.66 & 0.95 & 1.61 & 1.00 & 1.87 & 0.89 & 2.28 & 0.88 & 
1.27 & 0.80 & 1.47 & 0.85 & 2.66 & 0.91 & 1.93 & 0.76 & 1.39 & 0.86 & 2.90 & 0.90 & 2.33 & 0.73 & 2.03 & 0.84 & 
4.46 & 0.92 & 2.58 & 0.71 & 1.95 \\ 
0.183 & 1.00 & 11.70 & 1.00 & 3.62 & 1.00 & 2.56 & 0.98 & 6.06 & 0.98 & 2.40 & 1.00 & 2.94 & 0.89 & 3.31 & 0.92 & 
1.98 & 0.85 & 2.45 & 0.85 & 3.95 & 0.91 & 2.96 & 0.81 & 2.49 & 0.87 & 4.53 & 0.91 & 3.96 & 0.77 & 3.67 & 0.84 & 
6.95 & 0.93 & 4.12 & 0.73 & 3.51 \\
0.209 & 1.00 & 4.26 & 1.00 & 1.70 & 1.00 & 1.38 & 1.00 & 3.05 & 0.99 & 1.81 & 1.00 & 1.80 & 0.95 & 2.15 & 0.92 & 
1.40 & 0.83 & 1.52 & 0.89 & 2.68 & 0.91 & 2.30 & 0.80 & 1.45 & 0.90 & 3.26 & 0.91 & 2.54 & 0.76 & 2.17 & 0.88 & 
5.21 & 0.93 & 2.72 & 0.73 & 2.03 \\
0.295 & 1.00 & 9.70 & 1.00 & 2.52 & 1.00 & 2.31 & 1.00 & 4.70 & 0.93 & 2.22 & 1.00 & 2.67 & 0.92 & 3.48 & 0.87 & 
1.82 & 0.80 & 2.06 & 0.88 & 4.06 & 0.88 & 2.79 & 0.76 & 1.97 & 0.89 & 4.31 & 0.88 & 3.49 & 0.73 & 3.03 & 0.86 & 
5.55 & 0.92 & 3.71 & 0.71 & 2.95 \\
0.314 & 1.00 & 8.46 & 1.00 & 2.10 & 1.00 & 1.64 & 1.00 & 4.43 & 0.95 & 1.58 & 1.00 & 2.03 & 0.94 & 2.58 & 0.89 & 
1.32 & 0.80 & 1.43 & 0.89 & 2.89 & 0.91 & 2.04 & 0.76 & 1.32 & 0.90 & 3.34 & 0.90 & 2.75 & 0.73 & 1.91 & 0.88 & 
4.94 & 0.93 & 2.77 & 0.69 & 1.94 \\
0.600 & 1.00 & 5.52 & 1.00 & 1.41 & 1.00 & 1.53 & 0.97 & 4.64 & 0.93 & 1.85 & 1.00 & 1.91 & 0.88 & 3.24 & 0.88 & 
1.45 & 0.80 & 1.54 & 0.84 & 3.54 & 0.88 & 2.42 & 0.77 & 1.51 & 0.86 & 4.05 & 0.89 & 2.91 & 0.73 & 2.24 & 0.83 & 
5.65 & 0.92 & 3.03 & 0.71 & 2.18 \\ 
0.620 & 1.00 & 6.48 & 0.99 & 2.00 & 1.00 & 1.47 & 0.99 & 3.29 & 0.91 & 1.91 & 1.00 & 1.90 & 0.90 & 2.62 & 0.86 & 
1.55 & 0.79 & 1.59 & 0.86 & 3.26 & 0.87 & 2.62 & 0.76 & 1.52 & 0.87 & 3.89 & 0.86 & 2.96 & 0.71 & 2.32 & 0.85 & 
6.16 & 0.91 & 3.26 & 0.61 & 2.19 \\ 
0.654 & 1.00 & 3.90 & 0.99 & 1.69 & 1.00 & 1.56 & 0.97 & 2.99 & 0.91 & 1.91 & 1.00 & 1.94 & 0.88 & 2.02 & 0.86 & 
1.58 & 0.75 & 1.72 & 0.84 & 2.78 & 0.88 & 2.46 & 0.72 & 1.66 & 0.86 & 3.53 & 0.87 & 2.87 & 0.68 & 2.43 & 0.83 & 
5.38 & 0.91 & 2.93 & 0.61 & 2.23 \\
0.892 & 0.97 & 3.42 & 1.00 & 1.38 & 1.00 & 1.51 & 0.89 & 3.82 & 0.98 & 1.96 & 1.00 & 1.94 & 0.81 & 2.50 & 0.92 & 
1.61 & 0.83 & 1.75 & 0.78 & 2.86 & 0.91 & 2.58 & 0.80 & 1.65 & 0.80 & 3.28 & 0.91 & 2.81 & 0.76 & 2.47 & 0.78 & 
5.09 & 0.93 & 3.17 & 0.72 & 2.38 \\ 
0.914 & 1.00 & 4.55 & 1.00 & 2.44 & 1.00 & 1.80 & 0.96 & 4.22 & 0.99 & 2.56 & 1.00 & 2.31 & 0.87 & 3.06 & 0.92 & 
2.16 & 0.85 & 2.14 & 0.83 & 4.54 & 0.92 & 3.44 & 0.81 & 2.18 & 0.85 & 5.42 & 0.92 & 3.70 & 0.77 & 2.99 & 0.82 & 
9.03 & 0.94 & 4.30 & 0.73 & 2.97 \\
\enddata
\end{splitdeluxetable*}

\newpage

\begin{splitdeluxetable*}{c|cr|cc|cc|cc|cc|cc|cc|cc|ccBcc|cc|cc|cc|cc|cc|cc|cc|cc}
\tabletypesize{\footnotesize}
\tablecaption{$f_{spot}$ \& \textbf{$\chi^2_{\rm red}$} Values for K7V$_p$ Template 
Models\label{tab:wholefittingK7}}
\tablehead{
\colhead{Phase} & \multicolumn{6}{c}{K7V$_{p}$+M3V$_{s}$} & \multicolumn{6}{c}{K7V$_{p}$+M4V$_{s}$} & 
\multicolumn{6}{c}{K7V$_{p}$+M5V$_{s}$} & \multicolumn{6}{c}{K7V$_{p}$+M6V$_{s}$} &
\multicolumn{6}{c}{K7V$_{p}$+M7V$_{s}$} &
\multicolumn{6}{c}{K7V$_{p}$+M8V$_{s}$}\\
\colhead{}         & \twocolhead{$F_{0.8-1.35\ \micron}$} & \twocolhead{TiO}   & \twocolhead{FeH} & 
\twocolhead{$F_{0.8-1.35\ \micron}$} & \twocolhead{TiO}   & \twocolhead{FeH} & \twocolhead{$F_{0.8-1.35\ \micron}$}
& \twocolhead{TiO}   &  \twocolhead{FeH} & \twocolhead{$F_{0.8-1.35\ \micron}$} & \twocolhead{TiO}   &  
\twocolhead{FeH} & \twocolhead{$F_{0.8-1.35\ \micron}$} & \twocolhead{TiO}   & \twocolhead{FeH} & 
\twocolhead{$F_{0.8-1.35\ \micron}$} & \twocolhead{TiO} & \twocolhead{FeH} \\
\hline
\colhead{$\phi$}   & \colhead{$f_{spot}^{best}$}   & \colhead{$\chi^2_{\rm red}$}  & \colhead{$f_{spot}^{best}$}  
& \colhead{$\chi^2_{\rm red}$} & \colhead{$f_{spot}^{best}$}   & \colhead{$\chi^2_{\rm red}$}            & 
\colhead{$f_{spot}^{best}$}   & \colhead{$\chi^2_{\rm red}$} & \colhead{$f_{spot}^{best}$}   & 
\colhead{$\chi^2_{\rm red}$}            & \colhead{$f_{spot}^{best}$}   & \colhead{$\chi^2_{\rm red}$} & 
\colhead{$f_{spot}^{best}$} & \colhead{$\chi^2_{\rm red}$}   & \colhead{$f_{spot}^{best}$}   & 
\colhead{$\chi^2_{\rm red}$}   & \colhead{$f_{spot}^{best}$} & \colhead{$\chi^2_{\rm red}$} & 
\colhead{$f_{spot}^{best}$}   & \colhead{$\chi^2_{\rm red}$}            & \colhead{$f_{spot}^{best}$}   & 
\colhead{$\chi^2_{\rm red}$} & \colhead{$f_{spot}^{best}$} & \colhead{$\chi^2_{\rm red}$}   & 
\colhead{$f_{spot}^{best}$}   & \colhead{$\chi^2_{\rm red}$}   & \colhead{$f_{spot}^{best}$} & 
\colhead{$\chi^2_{\rm red}$} & \colhead{$f_{spot}^{best}$}   & \colhead{$\chi^2_{\rm red}$}            & 
\colhead{$f_{spot}^{best}$}   & \colhead{$\chi^2_{\rm red}$} & \colhead{$f_{spot}^{best}$} & \colhead{$\chi^2_{\rm 
red}$}   & \colhead{$f_{spot}^{best}$} & \colhead{$\chi^2_{\rm red}$}
}
\decimals
\startdata
0.000 & 1.00 & 4.37	& 1.00 & 1.40 & 1.00 & 1.56 & 0.96 & 3.65 & 0.92 & 1.80 & 1.00 & 2.02 & 0.86 & 2.68 & 0.86 & 
1.54 & 0.68 & 1.67 & 0.82 & 3.28 & 0.88 & 2.14 & 0.65 & 1.53 & 0.83 & 4.23 & 0.87 & 2.57 & 0.62 & 2.19 & 0.81 & 
6.56 & 0.91 & 2.75 & 0.59 & 2.06 \\
0.028 & 1.00 & 3.65 & 1.00 & 1.43 & 0.99 & 1.17 & 0.95 & 3.31 & 0.93 & 1.50 & 1.00 & 1.53 & 0.85 & 2.45 & 0.87 & 
1.43 & 0.67 & 1.40 & 0.81 & 2.98 & 0.88 & 2.01 & 0.62 & 1.39 & 0.83 & 3.71 & 0.88 & 2.50 & 0.59 & 1.99 & 0.80 & 
5.70 & 0.92 & 2.61 & 0.56 & 2.08 \\ 
0.054 & 1.00 & 4.89	& 1.00 & 1.70 & 1.00 & 1.49 & 0.97 & 2.71 & 0.95 & 1.75 & 1.00 & 1.87 & 0.87 & 2.51 & 0.88 & 
1.60 & 0.71 & 1.54 & 0.82 & 2.98 & 0.91 & 2.09 & 0.67 & 1.50 & 0.84 & 3.31 & 0.90 & 2.51 & 0.62 & 2.11 & 0.81 & 
5.25 & 0.92 & 2.69 & 0.59 & 2.01 \\ 
0.183 & 1.00 & 11.70 & 1.00 & 2.52 & 1.00 & 2.56 & 0.98 & 6.06 & 0.98 & 2.19 & 1.00 & 2.94 & 0.88 & 3.46 & 0.92 & 
1.89 & 0.79 & 2.57 & 0.83 & 4.45 & 0.91 & 2.79 & 0.73 & 2.54 & 0.85 & 5.19 & 0.91 & 3.43 & 0.69 & 3.73 & 0.82 & 
8.31 & 0.93 & 3.65 & 0.65 & 3.56\\ 
0.209 & 1.00 & 4.26 & 1.00 & 2.10 & 1.00 & 1.38 & 1.00 & 3.13 & 0.99 & 1.56 & 1.00 & 1.80 & 0.93 & 2.53 & 0.92 & 
1.30 & 0.77 & 1.60 & 0.87 & 3.21 & 0.92 & 1.99 & 0.72 & 1.57 & 0.89 & 3.96 & 0.92 & 2.43 & 0.68 & 2.20 & 0.86 & 
6.34 & 0.94 & 2.59 & 0.65 & 2.10 \\
0.295 & 1.00 & 9.70	& 1.00 & 1.41 & 1.00 & 2.31 & 1.00 & 4.70 & 0.98 & 1.92 & 1.00 & 2.67 & 0.91 & 3.78 & 0.92 & 
1.53 & 0.71 & 2.14 & 0.85 & 4.42 & 0.91 & 2.13 & 0.67 & 2.07 & 0.87 & 4.81 & 0.91 & 2.78 & 0.62 & 3.11 & 0.84 & 
6.81 & 0.93 & 2.96 & 0.59 & 2.93 \\
0.314 & 1.00 & 8.46 & 1.00 & 2.00 & 1.00 & 1.64 & 1.00 & 4.43 & 0.99 & 1.91 & 1.00 & 2.03 & 0.93 & 2.65 & 0.92 & 
1.84 & 0.68 & 1.42 & 0.87 & 3.16 & 0.91 & 2.65 & 0.65 & 1.35 & 0.89 & 3.72 & 0.91 & 3.07 & 0.61 & 1.95 & 0.86 & 
5.68 & 0.93 & 3.38 & 0.58 & 1.91\\
0.600 & 1.00 & 5.52	& 1.00 & 1.69 & 1.00 & 1.53 & 0.96 & 4.85 & 0.93 & 1.93 & 1.00 & 1.91 & 0.86 & 3.60 & 0.87 & 
1.79 & 0.73 & 1.65 & 0.81 & 4.10 & 0.88 & 2.34 & 0.68 & 1.71 & 0.83 & 4.74 & 0.88 & 2.90 & 0.64 & 2.38 & 0.81 & 
6.91 & 0.92 & 2.99 & 0.60 & 2.22 \\
0.620 & 1.00 & 6.48	& 1.00 & 1.39 & 0.98 & 1.47 & 0.99 & 3.34 & 0.95 & 1.84 & 1.00 & 1.90 & 0.89 & 2.97 & 0.89 & 
1.56 & 0.69 & 1.69 & 0.84 & 3.82 & 0.91 & 2.26 & 0.64 & 1.70 & 0.85 & 4.70 & 0.90 & 2.68 & 0.60 & 2.44 & 0.82 & 
7.62 & 0.93 & 3.05 & 0.59 & 2.28 \\
0.654 & 1.00 & 3.90	& 1.00 & 2.44 & 0.96 & 1.56 & 0.97 & 2.90 & 0.93 & 2.69 & 1.00 & 1.90 & 0.86 & 2.34 & 0.88 & 
2.57 & 0.63 & 1.87 & 0.82 & 3.21 & 0.88 & 3.39 & 0.60 & 1.90 & 0.84 & 4.37 & 0.89 & 3.96 & 0.54 & 2.54 & 0.81 & 
6.88 & 0.92 & 4.38 & 0.51 & 2.36 \\ 
0.892 & 0.96 & 3.44 & 0.99 & 1.75 & 1.00 & 1.52 & 0.88 & 3.56 & 0.91 & 1.54 & 1.00 & 1.94 & 0.78 & 2.88 & 0.86 & 
1.22 & 0.77 & 1.87 & 0.75 & 3.51 & 0.87 & 1.82 & 0.72 & 1.80 & 0.78 & 4.17 & 0.86 & 2.24 & 0.66 & 2.52 & 0.75 & 
6.08 & 0.91 & 2.53 & 0.64 & 2.60 \\ 
0.914 & 1.00 & 4.55	& 0.99 & 3.62 & 1.00 & 1.84 & 0.95 & 4.30 & 0.91 & 2.43 & 1.00 & 2.31 & 0.85 & 3.67 & 0.86 & 
2.27 & 0.80 & 2.30 & 0.80 & 5.21 & 0.88 & 3.13 & 0.73 & 2.27 & 0.82 & 7.15 & 0.87 & 3.96 & 0.69 & 3.09 & 0.80 & 
10.48 & 0.91 & 4.32 & 0.65 & 3.01 \\
\enddata
\end{splitdeluxetable*}





\end{document}